# Incentive Mechanisms for Participatory Sensing: Survey and Research Challenges

FRANCESCO RESTUCCIA and SAJAL K. DAS, Missouri University of Science and Technology
JAMIE PAYTON, University of North Carolina at Charlotte

Participatory sensing is a powerful paradigm which takes advantage of smartphones to collect and analyze data beyond the scale of what was previously possible. Given that participatory sensing systems rely completely on the users' willingness to submit up-to-date and accurate information, it is paramount to effectively incentivize users' active and reliable participation. In this paper, we survey existing literature on incentive mechanisms for participatory sensing systems. In particular, we present a taxonomy of existing incentive mechanisms for participatory sensing systems, which are subsequently discussed in depth by comparing and contrasting different approaches. Finally, we discuss an agenda of open research challenges in incentivizing users in participatory sensing.

## 1. INTRODUCTION

Recent smartphones are equipped with a plethora of embedded multi-modal sensors, and integrate wireless communication technologies as well as complex processing capabilities. These technological features, coupled with the widespread usage of smartphones [Cisco 2014], have contributed to the emergence of applications based on a new and promising paradigm known as participatory sensing[1]. The main idea behind participatory sensing is to enable ordinary citizens (or users) to actively monitor various phenomena pertaining to themselves (e.g., health, social connections) or their community (e.g., environment). For example, the cameras on smartphones can be used as video and image sensors [Bao and Roy Choudhury 2010], the microphone can be used as an acoustic sensor [D'Hondt et al. 2013], and the embedded global positioning system (GPS) receiver can be used to gather accurate location information, while gyroscopes, accelerometers, and proximity sensors can be used to extract contextual information about the user (e.g., if the user is driving [Nawaz and Mascolo 2014]). Further, additional sensors can be easily interfaced with the phone via Bluetooth or wired connections (e.g., temperature sensors [Thermodo 2014]). Real-life applications, which can take advantage of both low-level sensor data and high-level user activities, range from real-time traffic monitoring applications like *Nericell* [Mohan et al. 2008] or *Waze* [Waze 2014] to air pollution monitoring [Méndez et al. 2011] and social networking [Miluzzo et al. 2007]. For an excellent survey of applications based on the participatory sensing paradigm, please refer to [Khan et al. 2013].

---

[1]For the sake of generality, in this paper we use the term *participatory sensing* to designate applications where participants voluntarily contribute sensor data for their own benefit and/or the benefit of the community by using their phones. Such a notion therefore includes mobiscopes [Abdelzaher et al. 2007], opportunistic sensing [Campbell et al. 2006], and equivalent terms such as mobile phone sensing, smartphone sensing [Lane et al. 2010], or crowdsensing. It also covers specific terminologies focusing on particular monitoring subjects, such as urban sensing [Campbell et al. 2006], citizen sensing [Burke et al. 2006], people-centric sensing [Campbell et al. 2006], [Campbell et al. 2008], and community sensing [Krause et al. 2008].

Although a clear consensus on the best architecture for participatory sensing systems has not been reached yet, the majority of the existing participatory sensing applications utilize a centralized cloud-based architecture. In particular, volunteers use mobile phones to collect sensor data and submit via wireless data communication links to a participatory sensing platform (PSP) located in the cloud. The sensing tasks on the phones can be triggered manually, automatically, or based on the current context [Christin et al. 2011]. On the PSP, the data is analyzed and made available in various forms, for example, graphical representations or maps showing the sensing results at an individual or community scale. The results may be displayed locally on the users' mobile phones or accessed by the broader public through web-portals, depending on the application needs.

Participatory sensing provides a significant number of advantages with respect to previous sensing paradigms:

— The lack of a fixed sensing infrastructure dramatically eases deployment and maintenance costs associated with the administration of the participatory sensing system;
— The sheer number of smartphone users, coupled with the ubiquitousness of WiFi and 3/4G cellular-based Internet connectivity, allows a level of spatio-temporal sensing coverage impossible to achieve in previous paradigms, including wireless sensor networks;
— The presence of people in the sensing loop provides the opportunity to acquire *opinions* along with sensor data, allowing the emergence of complex mobile applications such as real-time traffic monitoring [Thiagarajan et al. 2009], [Waze 2014], [Zhou et al. 2014];
— The widespread availability of software development tools and markets for smartphone applications (apps) makes development and distribution of participatory sensing software relatively easy.

The most distinctive characteristic of participatory sensing is that it relies completely on the voluntary commitment of participants to submit up-to-date and reliable information to the PSP. This implies that a key factor for the success of participatory sensing applications is incentivizing users' active participation in the sensing surveys. To this end, a number of incentive mechanisms, mostly based on game theory [Tadelis 2013] and auction theory [Krishna 2009], have been proposed to increase not only the collected amount of data, but also its quality. The underlying idea of such incentive mechanisms is to reward users depending on their contribution to the participatory sensing campaign. For example, rewards made may be based on submitting a report close to a desirable location [Jaimes et al. 2012], how a report contributes to the social welfare [Luo and Tham 2012; Tham and Luo 2014], the number of reports that a user sent, or the time dedicated to collecting and submitting sensing reports [Yang et al. 2012].

A number of research issues regarding incentivizing users in participatory sensing systems, however, still remain. For example, it is unclear what is the best way to incentivize users in participatory sensing. In particular, while some people may respond to a monetary reward to participate, others may still be willing to participate for free, provided they receive sufficient value by gaining access to the crowdsensed data that is shared via the participatory sensing system. Thus, understanding the reasons that drive individuals to participate in participatory sensing campaigns could lead to designing more effective incentive mechanisms. Another challenge to overcome is the fact that participatory sensing apps may access particularly sensitive information, such as the visited locations of the users, as well as personal multimedia data such as photos and video. This may significantly discourage participation to the sensing campaigns, regardless of the amount of incentive put forth [Albers et al. 2013], [Krontiris et al. 2010]. Furthermore, most of the existing incentive mechanisms reward users irrespective of the quality of the information contained in the reports, which may undermine information reliability and encourage abuse of the system by malicious participants [Restuccia and Das 2014]. Therefore, designing mechanisms able to estimate and enforce information quality becomes of primary importance. However, enforcing quality of information in participatory sensing is extremely challenging, due to the possible presence of unreliable humans in the sensing loop. These and other research challenges still need to be analyzed and discussed—and this motivates our work.

Within the scope of this manuscript, we analyze the state-of-the-art in incentive mechanisms as applied to participatory sensing campaigns; besides describing the solutions currently applied, we also highlight their limitations, and discuss open issues as well as their impact on participation of users in sensing surveys. Specifically, our contributions can be summarized as follows:

— We propose a taxonomy of existing incentive mechanisms for participatory sensing systems.
— For each component of the taxonomy, we discuss and analyze existing incentive mechanisms.
— Based on this analysis, we identify and discuss an agenda of open research challenges that must be addressed to support widespread participation of users in sensing campaigns.

These contributions are presented as follows. Section 2 briefly describes the concept of participatory sensing. In Section 3, we introduce a taxonomy of existing incentivization mechanisms for participatory sensing. Sections 4, 5 and 6 presents an overview and an analysis of existing incentives as they relate to the organizational structure of the taxonomy. In Section 7, we discuss open research challenges, before concluding the paper in Section 8 with directions of future work.

## 2. WHAT IS PARTICIPATORY SENSING?

Participatory sensing [Burke et al. 2006] is a novel paradigm that leverages the widespread usage of smartphones to acquire up-to-date and fine-grained information about a location (or event) of interest. As depicted in Figure 1, most of the existing participatory sensing systems are generally organized into an architecture where the main components are the *users*, the *sensing application* (app), and the *participatory sensing platform* (PSP).

— The main component of the system is the *users*, whose task is to use their smartphones (and in particular, the sensing application) to capture different kinds of sensor data, such as location, images, sound samples, accelerometer data, biometric data, and barometric pressure. However, *opinions* about the sensing area or an environmental phenomenon, for example, traffic status or weather information, may also be provided by the users. In practical implementations of participatory sensing systems, users typically register with the system by providing using a username and password [Waze 2014] that allows their contributions to be uniquely identified.

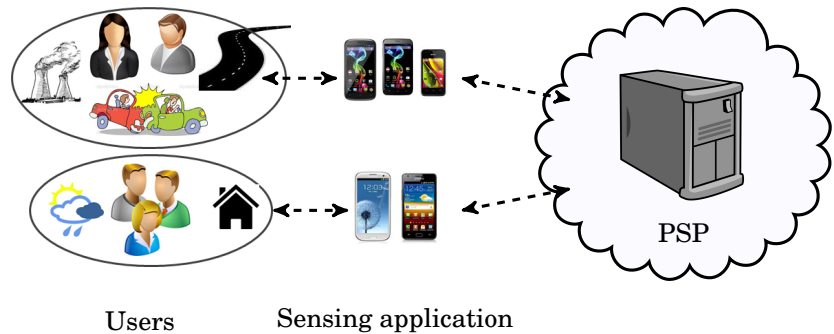

Fig. 1. Participatory sensing architecture.

— The *sensing application* (app), deployed on the users' smartphones, is distributed through common application markets like *Google Play* or *App Store*, or is retrieved from a mobile cloud computing system [Fakoor et al. 2012]. It is responsible for providing the users with a friendly user interface for data acquisition and visualization. In particular, data acquisition may be triggered by the users themselves or may be elicited by the app, on a one-time on-demand basis or periodically.

— The backend component of the system is the *participatory sensing platform* (PSP), responsible for the filtering, elaboration, and redistribution of sensed data, as well as coordinating every operation performed by the system. This component is usually implemented by a set of servers dedicated to the processing of sensed data [Restuccia and Das 2014]. The PSP also ensures efficient storage and elaboration of the sensed data coming from the users, which may be stored in relational databases [Gaonkar et al. 2008], or databases specially adapted to the management of sensor readings, for example, *sensedDB* [Grosky et al. 2007]. Along with the main elaboration system, the PSP might leverage a *reputation system* and an *incentive mechanism*. Briefly, the target of a reputation system is to predict the reliability of the data sent by the users based on their past behavior [Wang et al. 2013; Huang et al. 2010] so as to filter out unreliable reports. Conversely, the target of an incentive mechanism is to encourage participation of users by appropriately rewarding the users for their contributions to the participatory sensing campaign. The detailed discussion of such incentive mechanisms is the focus of this paper.

## 3. OVERVIEW OF INCENTIVE MECHANISMS IN PARTICIPATORY SENSING

In this section, we first identify the main components of the sensing and incentivization process of existing systems. We then propose a taxonomy of incentive mechanisms for participatory sensing, which will be used to organize a detailed discussion of the state of the art in the following sections.

Figure 2 depicts the main activities of the sensing and incentivization processes in a participatory sensing system. Since this collection of activities is performed again and again over the lifetime of the participatory sensing campaign, henceforth, we will use the term *sensing round* to refer to the execution of the following four steps.

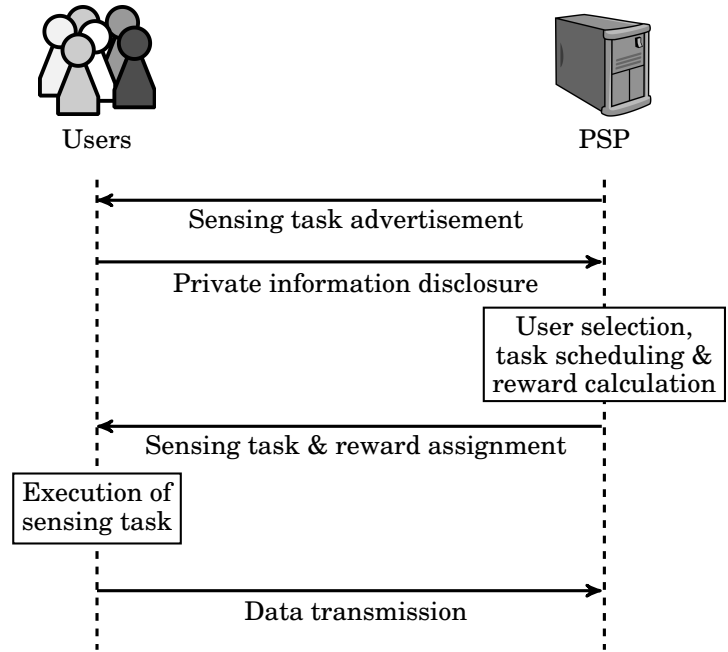

Fig. 2. Sensing and incentivization activities within a sensing round.

(1) *Sensing task advertisement.* In this phase, the participatory sensing platform (PSP) communicates to the users the list of sensing tasks that need to be executed during the current sensing round. In particular, each sensing task specifies a series of requirements, such as the sampling rate requested [Koutsopoulos 2013], minimum sensing time [Ji and Chen 2014], maximum distance from specified location [Ueyama et al. 2014], or task expiration time [Li and Cao 2013; Zhao et al. 2014]. For example, a sensing task might be "report the current traffic status near the Golden Gate bridge by 5:00PM". Additional parameters may be sent, such as quality of information requirements [Albers et al. 2013; Liu et al. 2011]. Sensing tasks can be advertised statically [Yang et al. 2012; Feng et al. 2014a; Duan et al. 2014] or dynamically [Restuccia and Das 2014; Feng et al. 2014b]. In some cases, depending on the application, tasks can be retrieved by the users asynchronously, e.g., each day [Li and Cao 2014], or whenever requested by the participants [Li and Cao 2013].

(2) *Private information disclousure.* After the advertisement of the sensing tasks, the PSP collects information about the participating user, often called the *type* [Yang et al. 2012] of the user, which can be leveraged by the PSP to make a choice regarding the scheduling of sensing tasks. For example, users may supply a *bidding value*, which is included in the user type for use in auction-based incentive mechanisms. The PSP may collect temporal information about the user, which can be used to determine the user's availability to perform sensing services [Han et al. 2013]. The PSP may also acquire information regarding the characterization of the skills of a person [Luo et al. 2014c; Luo et al. 2014b] and the incurred cost for the required sensing services, for example, privacy loss [Duan et al. 2012], energy consumption, or mobile data charges [Luo and Tham 2012].

(3) *User selection, task scheduling, reward calculation and assignment.* After receiving private information describing the users, the PSP selects a subset of users that will submit the sensed information to the PSP, and schedules the sensing tasks for users. For example, users might be selected according to their geographical position [Ma et al. 2014; Feng et al. 2014a; Zhang et al. 2014], or according to the submitted bid [Yang et al. 2012], cost of sensing services [Duan et al. 2014], or sensing effort [Luo et al. 2014b]. Furthermore, the PSP may schedule the sensing tasks according to the temporal availability of the users [Chen and Wang 2013], [Feng et al. 2014b]. In this phase, the users are also rewarded for their services, usually according to the time dedicated to the sensing services. The information quality contained in sensing reports might also be a parameter to decide the amount of reward to assign to users [Restuccia and Das 2014]. User selection, task scheduling, and reward assignment algorithms are an important part of incentive mechanisms, and are a focus of our analysis of the state of the art in Sections 4, 5, and 6.

(4) *Execution of sensing task and data transmission.* After being selected and instructed on the sensing task to execute, a user is allowed to begin performing the sensing service using the sensing application. The sensing application may be designed to assist in collecting data according to one of the following sensing modes: manual, automatic, or context-aware [Christin et al. 2011]. In the *manual* mode, the participants explicitly trigger the collection of sensor readings when they detect what they perceive to be relevant events, such as traffic congestion. On the contrary, the participants are not directly involved in the *automatic* mode (also known as opportunistic sensing [Khan et al. 2013]), where a sensing application continuously runs in the background and regularly sends the sensed information to the PSP, as for example in *Nericell* [Mohan et al. 2008]). Data collection might also be triggered by leveraging the *context*, occurring upon the detection of an event or condition (e.g., clapping of hands, as in [Bao and Roy Choudhury 2010]). The sensing application handles the transfer of the sensed data from the smartphone to the PSP, making use of communication infrastructure available to the mobile phone, such as WiFi or 3G/4G connectivity. For example, the sensor readings can be transmitted to the server using SMS or TCP connections [Gaonkar et al. 2008], remote procedure calls [Miluzzo et al. 2007], or web interfaces [Fakoor et al. 2012]. Recently, opportunitistic forwarding coupled with data fusion has been proposed as a viable method for data transmission [Ma et al. 2014].

### 3.1. Taxonomy of incentive mechanisms

Figure 3 illustrates the proposed taxonomy of the approaches used for incentivizing users in participatory sensing. The taxonomy first differentiates existing work based on the *purpose* of the mechanisms; specifically, the first distinction is made between *application-specific* and *general-purpose* incentive mechanisms. This distinction is necessary since the former are ad-hoc mechanisms proposed as a part of the participatory sensing system, and are tailored to incentivize users' participation to the specific application only; in other words, they are not designed in a way that is generalizable for broader application. On the other hand, the latter are mechanisms which do not assume any particular participatory sensing application, but instead are general in purpose and can be applied to a wide spectrum of systems satisfying the assumptions of the mechanism. As yet, the majority of the mechanisms proposed have been in the general-purpose category.

Among the general-purpose mechanisms, another distinction is made based on the *methodology* being used to incentive participation. In particular, we differentiate between *game-theoretical* approaches, which are grounded in mathematical models that provide a foundation for reasoning about rational decision-making, and mechanisms that exploit different concepts, such as micropayments [Reddy et al. 2010b]. Game-theoretical approaches can be further categorized, distinguishing between mechanisms that exploit auction theory [Krishna 2009] and mechanisms that instead use other game-theoretical concepts such as Stackelberg games [Tadelis 2013].

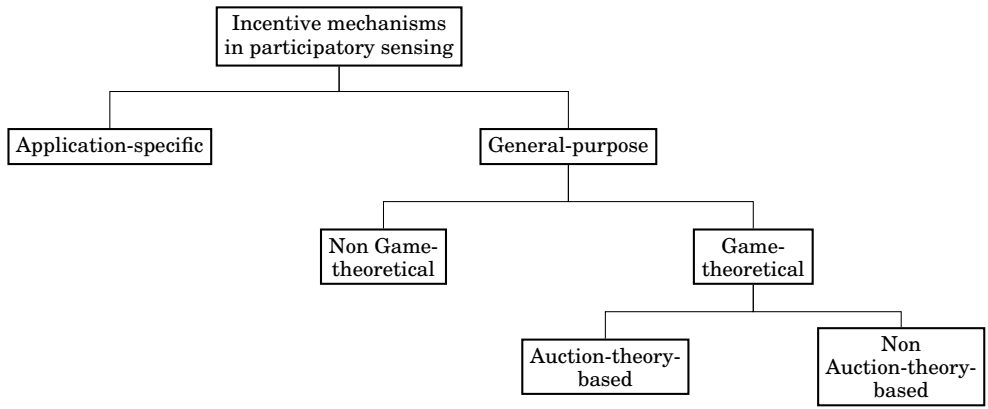

Fig. 3. Taxonomy of existing incentive mechanisms.

In the following sections, we extensively discuss the most significant systems proposed so far in each category of the taxonomy. For each category, we will characterize the foundations of the class of participatory sensing incentive systems, review the state of the art to highlight key differentiating ideas, and analyze approaches, discussing advantages and limitations of key elements of individual approaches and, more broadly, the class of approaches.

## 4. APPLICATION-SPECIFIC INCENTIVE MECHANISMS

Although many participatory sensing applications have been proposed so far [Khan et al. 2013], the problem of incentivizing users' participation has been overshadowed by other design challenges, such as processing sensed data [Lane et al. 2010]. Indeed, we identified as few as three systems that include an incentivization mechanism as a first-class element of their design.

One such system is *LiveCompare* [Deng and Cox 2009], which is designed to improve interstore grocery price comparisons for shoppers. *LiveCompare* addresses the challenge of user incentivization through a simple query protocol that results in a *quid pro quo* interaction. When a user at a grocery store wants a price comparison for a product of interest, he uses the *LiveCompare* app installed on his mobile phone to identify the product by snapping a photograph of the product's price tag; the photo is then provided to the *LiveCompare* PSP as a price comparison query. The product of interest is uniquely identified via a barcode included on the price tag. In exchange for submitting this price data point, the user receives pricing information for the scanned product at other nearby grocery stores. The photograph provided by the user is uploaded to a central repository for satisfying future queries.

Another system which embeds an incentive in the design is *CrowdPark* [Yan et al. 2011]. This system allows users to "loosely reserve" parking spots in urban areas, and is based on the interesting idea that an individual who is currently parked can provide a notification in advance about when she plans to leave. This information may be *sold* to a buyer who is willing to pay to reserve the parking spot. This simple protocol for drivers to buy and sell information about parking vacancies acts as an incen-tive mechanism for both buyers and sellers of parking information. The authors point out two interesting concepts when designing the *CrowdPark* incentive. In particular, they remark that the incentive mechanism should take into account the "utility" of a transaction - that is, bonuses should be provided when a transaction leads to successful acquisition of a parking spot, and refunds should be provided when unsuccessful. However, on the other hand, there is also need to handle *malicious users*, whose goal is solely to maximize their monetary gain, either by masquerading as fake sellers or by denying successful reservations to obtain a refund. To this end, the authors proposed a simple game-theoretical design of the parking reservation rules which ensures that buyers maximize their gain by telling the truth. The key idea behind the approach is that a buyer who successfully parks at a reserved spot can re-sell that spot through *CrowdPark* if they tell the truth. If they deny a successful instance of parking in a reserved spot, obviously they cannot re-sell this spot. Therefore, if reward parameters are set to ensure that the average gain of re-selling is higher than the gain by lying and receiving a refund, a rational user will be inclined towards telling the truth. The extended problem formulation was extended by the same authors in [Hoh et al. 2012], and analyzed by using Amazon Mechanical Turk (AMT) [Amazon 2014] as the experimental platform. They recruited 131 participants and designed a web-based survey that replicates the actual scenario of consumer confirmation. Results indicate that the proposed methodology drives users towards cooperative behavior.

To the best of our knowledge, the only other participatory sensing application to include an incentive mechanism as a key part of the system design is *Waze* [Waze 2014]. In particular, the reward strategy of Waze leverages the concept of *"gamification"* [Deterding et al. 2011] to engage users and encourage them to provide more information without the need for monetary incentives. For example, it allows drivers to "drive over" icons of cupcakes and other objects to earn points. Waze also uses a point-based system to rank users based on their level of participation. Specifically, Waze offers points for traffic or road hazard reports, which can be used to change the user's avatar, and to increase their status in the community. Users have their own personal "dashboard" where their ranking among worldwide users is shown. There is also a leaderboard, that shows who has accrued the most points over all, or who has driven the most miles this week, or who has filed the most reports [Furchgott 2010]. The concept of gamification as applied to participatory sensing has also been recently explored in [Ueyama et al. 2014], and will be detailed in Section 5 as a non-game-theoretical mechanism to incentivize user participation.

### 4.1. Discussion

The incentivization methods that have been proposed as part of the *LiveCompare* [Deng and Cox 2009] and *Crowdpark* [Yan et al. 2011] systems make strong assumptions that, ultimately, limit the applicability of the proposed mechanisms. For example, the effectiveness of the approach proposed in *LiveCompare* has not been validated through a systematic user study or mathematical analysis. In particular, since the study of the *LiveCompare* relied on a user survey through AMT rather than a deployment that required users to perform the sensing task [Deng and Cox 2009], it is unclear if a *quid pro quo* scheme in which users receive information from the PSP (e.g., receiving price information at nearby stores for a product of interest) is enough of an incentive for a user to perform high-burden sensing tasks (e.g., taking photos for a significant number of items while grocery shopping).

Furthermore, the *CrowdPark* approach, proposed in [Yan et al. 2011] and extended in [Hoh et al. 2012], assumes that mobile phones are trustworthy, and therefore a malicious user cannot manipulate location coordinates intentionally when reporting the location and availability of a parking space. However, it is well-known that users can easily spoof GPS coordinates by using widely available applications such as *FakeLocator*, available for both iOS and Android-based smartphones. This piece of software is able to change the current GPS position of users by hijacking the communication between the smartphone location interfaces and applications accessing the location data [Restuccia and Das 2014]. For example, *FakeLocator* acts as an additional location provider that takes a higher priority than all others, forcing the Android system to distribute the fake location that it is advertising. The user can then specify any location on Earth to *FakeLocator* and let the other apps in the smartphone "believe" that the user is in that location. Hereafter, we refer to this kind of malicious behavior as a *location-spoofing attack*. Figure 4 depicts the functioning scheme of *FakeLocator* and an example of a location-spoofing attack.

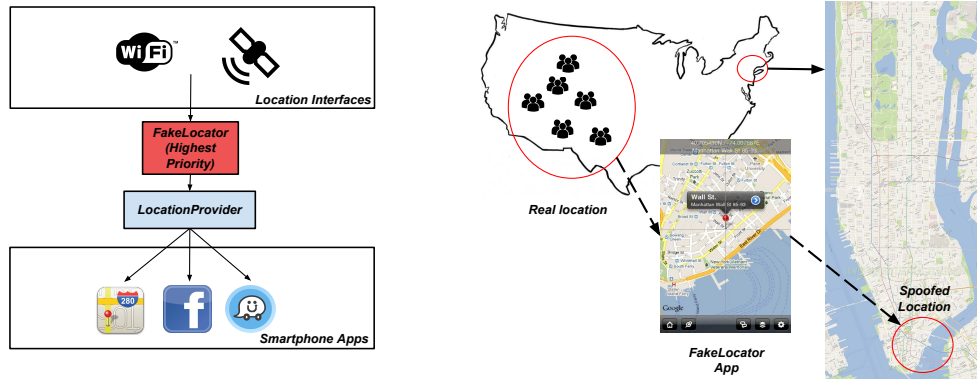

Fig. 4. Functioning scheme of *FakeLocator* and location-spoofing attack.

*Waze* has proven to be effective in incentivizing user participation, as more than 50 millions users participate daily as users of and contributors to the traffic map [Yahoo! 2012]. However, the gamification incentive used by *Waze* is also prone to the location-spoofing attack, as the system is not able to recognize between genuine and compromised users. This vulnerability of *Waze* was also demonstrated by a large-scale attack in which a fake traffic jam was injected into the system [Atherton 2014]. Therefore, the game-based incentive system can be compromised by malicious users, who can continue to earn pointsthrough the gamification incentive without actually performing any sensing tasksof value.

## 5. NON GAME-THEORETICAL INCENTIVE MECHANISMS

This section surveys general-purpose incentive mechanisms that do not exploit game-theoretical concepts in their design. Table I summarizes the papers (in alphabetical order) discussed in this section, with a brief description of the key contribution. In order to facilitate the discussion, we further divide the work presented in this section into three categories:

— *Quality of Information (QoI)-aware mechanisms*. These mechanisms take into account the QoI of each sensing report before making decisions on sensing task and reward assignment. Since QoI in participatory sensing systems has not been clearly defined yet, the characterization and assessment of QoI differ across the incentive mechanisms surveyed.
— *Resource-aware mechanisms*. These mechanisms consider the resource consumption (e.g., energy, mobile cellular data) of the user in assigning sensing tasks.

Table I. Summary of non game-theoretical-based incentive mechanisms.

| Paper | Key contribution | QoI | Resources | Privacy |
|---|---|---|---|---|
| [Cheung et al. 2014] | Resource-aware incentive mechanism | × | ✓ | × |
| [Kawajiri et al. 2014] | Gamification-based mechanism | ✓ | × | × |
| [Li and Cao 2014] | Two privacy-preserving incentive mechanisms | × | ✓ | ✓ |
| [Li and Cao 2013] | Privacy-aware incentive mechanism including multiple user contributions | × | × | ✓ |
| [Liu et al. 2011] and [Liu et al. 2014] | QoI-aware and energy-efficient incentive framework | ✓ | ✓ | × |
| [Luo et al. 2014a] | Endorsement-based incentive mechanisms through web relations | ✓ | × | × |
| [Mukherjee et al. 2014] | Incentive mechanism based on types and priorities of events, quality and timeliness of reports and user intent | ✓ | × | × |
| [Reddy et al. 2010a] | User selection based on spacial and temporal availability as well as participation habits | ✓ | × | × |
| [Reddy et al. 2010b] | Micropayments-based mechanism | ✓ | × | × |
| [Ren et al. 2014] | Social impact and trustworthiness of users | ✓ | × | × |
| [Restuccia and Das 2014] | Trust-based framework based on mobile security agents | ✓ | × | × |
| [Song et al. 2014] | QoI-aware incentive mechanism with budget constraints | ✓ | × | × |
| [Tham and Luo 2013] and [Tham and Luo 2015] | Incentive mechanism with novel definition of Quality of contributed Service and market equilibrium | ✓ | × | × |
| [Ueyama et al. 2014] | Gamification-based mechanism | × | × | × |
| [Zhang et al. 2014] | Recruitment framework with budget constraints | ✓ | × | × |

— *Privacy-aware mechanisms.* These mechanisms address the need to provide privacy to the participating users who are exposing sensitive data.

## 5.1. QoI-aware mechanisms

For participatory sensing applications to be used as effective tools for collecting data that will be interpreted into high-level information used in decision- making, it is essential to give careful consideration to the quality of information (QoI) that can be derived from participatory sensing data. Methods for assessing the quality of information (QoI) and encouraging participants to contribute data that meets QoI requirements for participatory sensing campaigns continues to emerge as an important research topic. Several definitions of QoI for participatory sensing have been put forth, including spatial coverage, temporal coverage, and timeliness. QoI metrics for participatory sensing also focus on the quality of the provider, assessesing the reputation of the participant for providing useful data as part of the participatory sensing campaign. Below, we discuss incentive mechanisms that are designed to improve the quality of information provided by users in participatory sensing systems.

*5.1.1. QoI by recruitment, micro-payments and reputation.* To improve the QoI achieved through participatory sensing, Reddy et al. proposed a recruitment process to select participants that are likely to provide information that meet QoI requirements [Reddy et al. 2010a]. In this work, QoI is defined in terms of the ability to collect a crowd-sensed data set that meets spatial and temporal coverage requirements. As such, the system considers availability predictions of volunteers for recruitment of participants and their reputations are introduced for incentivization. In detail, the system relies on three steps to select the users:

—*Qualifier:* Participants must meet minimum requirements, such as destinations and routes within time, space, and transportation mode constraints.
—*Assessment:* Once participants that meet minimum requirements are found, the recruitment system then identifies which subset of individuals maximize coverage over a specific area and time period while adhering to the required transportation modes.
—*Progress Review:* As a campaign runs, the recruitment system must check participants' coverage and data collection reputation to determine if they are consistent with their profile.

To build the reputation score of users, the authors use sampling likelihood, as well as assessments of quality and validity of a user's contributions over several campaigns or by campaign-specific calibration exercises. In particular, they use a Beta function-based algorithm to calculate the reputation, using the the correct and incorrect sensing reports as $\alpha$ and $\beta$ parameters, respectively. The authors demonstrate the application of the framework by analyzing data from a pilot mobility study consisting of ten users.

Similarly, in [Zhang et al. 2014] the authors propose *CrowdRecruiter*, a framework that aims to minimize incentive payments by selecting a small number of participants while satisfying a QoI requirement framed as a *probabilistic* spatiotemporal coverage constraint. To reduce the barrier to participation, CrowdRecruiter leverages piggybacking on 3G phone calls to send data in an effort to minimize energy consumption on the users' mobile devices. Information about the 3G connection is also used to determine how to satisfy the QoI requirement. First, the framework predicts the call and coverage probability of each mobile user based on historical records. It then efficiently computes the joint coverage probability of multiple users as a combined set and selects the near-minimal set of participants, which meets the coverage ratio requirement in each sensing cycle.

The concept of *micro-payments*, in which sensing tasks are matched with small payments, has been also proposed in [Reddy et al. 2010b] as an incentive model to increase high-quality contributions from users. As the first such effort, contributions of this work include the definition of a set of metrics that can be used to evaluate the effectiveness of an incentive to motivate an individual user. Ths set of metrics included the quantity of submissions, an application-specific measure of the quality of data submitted, and the spatial and temporal coverage. An evaluation of the approach was performed through a pilot study using various micro-payment schemes in a participatory sensing application for a university initiative; users were asked to take photos of trash bins on campus with the goal of understanding recycling habits across the campus. Results indicated that micro-payments were effective than a single lump payment for increasing quantity of submissions, application-specific quality of data (here, clear and focused images of trash bins located on the campus), and spatiotemporal diversity of submissions.

Because participatory sensing relies on the contributions of users, researchers have pointed out the potential for developing systems that identify *reliable* users [Huang et al. 2014] as a way to address QoI requirements. Although several reputation frameworks have been proposed (e.g., [Christin et al. 2013; Wang et al. 2014]), the authors of [Restuccia and Das 2014] point out that previous incentive and reputation mechanisms lack a scalable and secure method for validating the QoI of sensing reports. For this reason, they propose a novel trust-based data verification system based on the concept of *mobile security agents* (MSAs). In detail, MSAs are defined as users that periodically report information about their surroundings to the PS platform, and considered to be *reliable*, which means the information sent by MSAs is correct and reflects the real condition of the area nearby the MSA.

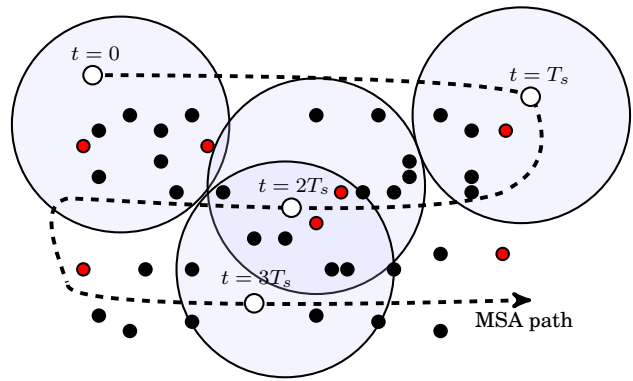

Fig. 5. An MSA roaming over the sensing area [Restuccia and Das 2014].

Figure 5 shows a sensing scenario in which only one MSA is roaming the sensing area. Legitimate and malicious users are depicted by black and red dots, respectively. Figure 5 also depicts the different locations over time of the MSA at different sensing rounds. At each sensing round, the MSA reports her sensed data to the PS platform (PSP). Then, such information is used to update the reputation of the users nearby the MSA at that particular time. In particular, the reputation is updated only for the users inside a portion of sensing area, defined as *update area* and depicted as a blue circle in Figure 5. The authors argue that this approach provides benefits such as scalability and security from location-based attacks as the location of users is not involved in computing their reputation.

*5.1.2. QoI by gamification.* Exploiting *gamification* [Deterding et al. 2011] has also been recently proposed as a viable and effective technique to incentivize users' participation and also increase QoI. Briefly, gamification is the use of game thinking and mechanics in non-game contexts to engage users in solving problems and increase users' self contributions. Widely known examples of systems that use gamification to encourage participation are the crowdsourced programming advice platform *Stack Overflow*[2], which rewards users who provide top-ranked answers; the job recruitment platform *HackerRank*[3], which presents users with coding "challenges" that relate to open positions for which employers are seeking candidates; and the participatory sensing system Waze [Waze 2014], which uses several gamification elements, including leaderboards, as incentives for providing traffic information.

With respect to participatory sensing incentives that use gamification to address QoI requirements, Kawajiri *et al.* proposed in [Kawajiri et al. 2014] a framework called *Steered Crowdsensing*. The system displays points to users to motivate them to move to a specific location, which increases QoI as the spacial coverage of the sensing area is increased. The authors formalize the calculation of points paid to users by introducing a quality indicator of service in an online machine learning setting. A strong contribution of the paper is the deployment of a crowdsensing system to test the framework. A similar approach was also proposed in [Mukherjee et al. 2014], where different incentives were managed based on types and priorities of events, quality and timeliness of event reports as well as resident intent is used to incentivize users.

---

[2]http://www.stackoverflow.com
[3]http://www.hackerrank.com

Another approach that uses gamification to address QoI requirements is [Ueyama et al. 2014], where the authors proposed an incentive mechanism that incorporates a status level scheme depending on earned reward points, so that users with higher status can earn more reward points. They also introduced a ranking scheme among users and a badge scheme so that users are attracted by getting not only monetary reward points but also by engendering a sense of accomplishment. They then formulate the problem of sensing given points of interest (PoI) with minimal reward points and devise a heuristic algorithm for deriving the set of users to which requests are sent and appropriate reward points for each request. The algorithm requires the participation probability distribution for each tuple of a user, reward points, and the burden of a sensing task. Similar to [Kawajiri et al. 2014], they performed an experiment with 18 users for 30 days; results indicate that the mechanism is effective in increasing the participation of users.

*5.1.3. Miscellaneous approaches.* One of the first research efforts to include QoI in the design of the incentive mechanism was [Liu et al. 2011] (extended in [Liu et al. 2014]), which presented a QoI- and energy-efficient incentive mechanism framework that maximizes the quality of sensed data with minimization of the incentive cost and energy consumption of smartphones. This is accomplished by first defining the *QoI satisfaction index*, which measures how close the received QoI is to the expected/desired quality. Then, by arguing that the reward must be assigned according to the expectation of users, they introduce in an analogous manner the *credit satisfaction index*, which measures the degree of credit satisfaction for individual smartphone users. These quantities are used as input to a mathematical model based on a *Gur Game* [Tung and Kleinrock 1993], where reward and energy loss guide the users' choice between QoI and energy-efficiency.

The game proposed is represented by a nearest-neighbor discrete-time Markov chain (DTMC) of size $2M$, where the set of states are divided in "positive" (from 1 to $M$) and "negative" states (from $-M$ to -1). Depending on the credit and QoI satisfaction indices at time step $j$, the state of the DTMC is shifted towards the rightmost positive state $M$, which represents the maximum possible level of contribution, or towards the leftmost state $-M$. If the current state at time step $j$ is positive (i.e., $j$ is between 1 and $M$), then the user chooses to participate, otherwise he does not. In such a way, the user autonomously chooses to participate or not depending on the desired tradeoff between personal satisfaction and energy consumption.

Recently, Tham *et al.* also explored the problem of incentivization with QoI in [Tham and Luo 2013] and [Tham and Luo 2015]. They first propose a new metric called *Quality of Contributed Service* (QCS) which characterizes the information quality and timeliness of a specific real-time sensed quantity achieved. The QCS is a cumulative time-decaying or timeliness-weighted quality of contribution: the more contributions, the higher the quality of the contributions, or the more up-to-date the contributions are in the consumable window, the higher the QCS value will be. Tham *et al.* also formulate a market-based framework with plausible models of the market participants comprising data contributors, service consumers, and a service provider. They analyze the market equilibrium under different conditions and obtain a closed form expression for the resulting QCS at market equilibrium. Some interesting theoretical findings are also shown, such as, optimal contribution level and rate for maximizing participant's payoff. An algorithm is also proposed to achieve market equilibrium.

Another interesting QoI-aware mechanism was recently proposed in [Song et al. 2014]. In this paper, the target is to find a subset of participants whose sensor data collection can best satisfy QoI requirements of multiple concurrent tasks in both temporal and spatial dimensions, by supposing a constrained budget for user reward. In particular, by defining as $u_i(\mathcal{X})$ the QoI satisfaction index (definition similar to [Liu et al. 2011]) obtained for the task $i \in \mathcal{Q}$ by a subset $\mathcal{X}$ of the set $\mathcal{M}$ of users, and by $d_m$ the reward demanded by user $m$, the problem is defined as follows:

$$\begin{aligned} \text{Maximize:} \quad & \overline{u}(\mathcal{X}) = \{u_1(\mathcal{X}), u_2(\mathcal{X}), \ldots, u_Q(\mathcal{X})\} \\ \text{subject to:} \quad & \mathcal{X} \subseteq \mathcal{M} \text{ and } \sum_{m \in \mathcal{M}} d_m \leq C_{tot} \end{aligned} \quad (1)$$

The authors argue that the optimization problem described in Equation 1 is multi-objective and therefore the optimal solution may not exist. Therefore, they propose a Pareto-optimal[4] version of the problem which is a modified non-linear Knapsack problem, known to be NP-complete. To solve the problem, the authors propose a dynamic participant selection strategy called DPS, where participants are selected based on a greedy algorithm that explicitly considers their expected amount of data collected, required QoI of multiple tasks, and users' incentive requirements, under the constraint of the aggregated task budget defined in Equation 1. To estimate the amount of data collected from each user, the authors introduce a Markov-based probability model based on the historical trajectory information of a user and the initial position. This information, joint with the sensing capabilities of the users, is used to predict the expected amount of data by all participants. The performance evaluation, conducted on real mobility traces, includes a comparison between DPS, a random selection algorithm and [Lee and Hoh 2010b]. Results indicate that DPS achieves better performance than both approaches.

### 5.2. Resource-aware mechanisms

A potential barrier to user contributions in participatory sensing campaigns is the consumption of resources associated with performing sensing tasks and providing sensing reports. Research in participatory sensing is beginning to address the need for resource-aware incentive mechanisms. The work in [Cheung et al. 2014] studies the decision process of the PSP and the users in participatory sensing applications that involve photo or video transmissions, where the reporting cost through the cellular network is non-negligible. This work differentiates from previous work in the fact that the proposed system model assumes Markovian[5] user mobility and WiFi availability only at predefined locations (i.e., connectivity is contrained). Data collection and user incentivization are modeled as a two-stage decision process, involving both the PSP and the users. In the first stage, the PSP announces the reward to the potential participants. Specifically, the authors propose a deadline-discounted reward scheme, where any participant who obtains measurement at its initial location $l_i(1)$ at time $t = 1$ and reports its data to the PSP at time $t$ will be given a reward $r = \theta^{t-1} \cdot R$, where $R \geq 0$ is the initial reward and $0 < \theta \leq 1$ is a discount factor. The PSP does not grant any reward for data reported after deadline $T$. In the second stage, each user makes its participation and reporting decisions to maximize its expected payoff, by considering its sensing cost, transmission cost, and the reward scheme of the SP. To this end, they propose an optimal participation and reporting decision (OPRD) algorithm, which achieves the maximal user payoff.

---

[4]A solution is Pareto-optimal if it is not possible to move from that solution and improve at least one objective function, without detriment to any other objective function.

[5]When the mobility is Markovian, time is divided into a set of time slots $\mathcal{T}$, while space is divided in a set of location $\mathcal{L}$. The current location $l_i(t) \in \mathcal{L}$ of user $i$ at time $t$ can be calculated by using the location transition matrix of user $i$, defined as $P_i \triangleq [p_i(l'|l)]_{L \times L}$, where $p_i(l'|l)$ is the probability that user $i$ will move to location $l'$ in the next time slot given that she is currently at location $l$.

### 5.3. Privacy-aware mechanisms

Privacy is a key concern that can potentially limit the widespread adoption of participatory sensing. The work by Li and Cao was the first approach to simultaneously deal with incentivization of users with privacy [Li and Cao 2013]. In this paper, they proposed two privacy-aware incentive schemes for participatory sensing to promote user participation. These schemes allow each mobile user to earn credits by contributing data without leaking which data she has contributed, and at the same time ensure that dishonest users cannot abuse the system to earn unlimited amount of credits. To achieve such goals, the basic idea was to issue one *request token* for each task to each user. The user consumes the token when she accepts the task. Since she does not have more tokens for the task, it cannot accept the task again. Similarly, to satisfy the second condition, each user will be given one *report token* for each task. When the PSP receives a report, it issues *pseudo-credits* to the reporting user, which can then be transformed into actual credits later on. To achieve the privacy goals, all tokens are constructed in a privacy-preserving way, such that a request (report) token cannot be linked to a user and credit tokens cannot be linked to the task and report from which the token was earned. The first scheme considers scenarios where a trusted third party (TTP) is available. It relies on the TTP to protect user privacy, and thus has very low computation and storage cost at each mobile user. The second scheme removes the assumption of TTP and applies blind signature and commitment techniques to protect user privacy. The ideas were extended by the same authors in [Li and Cao 2014], where the problem is extended to the more general case in which sensing tasks may require multiple reports from each user (e.g., in environmental monitoring applications).

### 5.4. Discussions

Although the problem of simultaneously incentivizing users and guaranteeing QoI has been recently tackled by a significant amount of research, the majority of existing approaches fail to consider the impact of users that *voluntarily* submit low-quality information. In particular, only [Restuccia and Das 2014] considers and tackles the location-spoofing attack described in Section 4.1. Furthermore, some assumptions in existing work oversimplify the actual complexity of estimating and enforcing information quality of sensing reports in participatory sensing systems. For example, in [Liu et al. 2011] and [Liu et al. 2014] the authors characterize the QoI as "the amount of information contributed by user $i$ corresponding to the $l$-th QoI attribute", which does not include any information directly related to quality, such as accuracy, timeliness of report, etc. Also, the proposed strategy based on Gur Game assumes that the user will always be rational and there will be no interaction of the user with the system, which is not entirely true in most of the existing participatory sensing applications. Similarly, in [Tham and Luo 2013] and [Tham and Luo 2015] the authors define QoI as "the detection accuracy of the system whose information quality is reflected in the degree of confidence that an event of interest has occurred" (i.e., yes/no decision). Although this assumption makes the analysis more tractable, it restricts the scope of the incentive framework only to applications handling binary opinions. The same problem is present in [Reddy et al. 2010a], where the very simple approach based on Beta function is unable to capture the complexity of sensing reports. The aspect of understanding and characterizing QoI is also ignored in [Restuccia and Das 2014], as the authors assume that an MSA can always postulate a binary decision (i.e., yes or no) regarding the QoI of a report. Finally, the system proposed in [Zhang et al. 2014] relies on piggybacking crowdsensing, therefore it is hardly applicable in reality as it assumes that 3G cellular data submitted by participants during phone calls is available to the participatory sensing application.

On the other hand, proposed gamification- and micropayments-based incentive mechanisms still lack systems to predict the quality of information that will be collected by users, and are based on heuristics rather than on strong mathematical principles. Also, strong assumptions on how QoI is determined undermine the applicability of the proposed mechanisms. For example, in [Kawajiri et al. 2014] the authors consider settings where sensory data can be continuously collected and analyzed, and sensory data arrives at regular time intervals, which is unrealistic in participatory sensing environments. As far as resource-aware systems are concerned, [Cheung et al. 2014] only considers transmission cost and network availability as resource parameters, and fails to consider the aspect of energy consumption in the analysis. Privacy-aware incentives proposed in [Li and Cao 2013] and [Li and Cao 2014], although technically sound, do not consider the aspect of QoI and resource-awareness in their design, therefore their applicability to real-world campaigns is limited. However, the most serious limitation of existing work on resource-aware and privacy-aware incentives is the lack of validation of the systems through actual participatory sensing campaigns. In particular, experimenting with real participants could allow us to understand the real value of privacy and resource consumption for participants, and therefore design more efficient and cost-effective mechanisms.

## 6. GAME-THEORETICAL INCENTIVE MECHANISMS

This section presents incentive mechanisms that include game-theoretical aspects in their design. We will discuss first auction-based approaches, and then we will focus on approaches that are based on other game theoretical foundations. For each of these categories, it is necessary to discuss first some preliminary concepts before analyzing approaches that exploit those foundations to incentive contributions to participatory sensing campaigns.

### 6.1. Preliminaries on Auction Theory

Auction theory is a branch of game theory that studies how the properties of *auctions* and how people act in auction markets. Auctions are decentralized market mechanisms for allocating resources [Krishna 2009], and are particularly interesting because they may be used to sell any item (i.e., auctions are *universal*), and the outcome of the auction does not depend on the identity of the bidders (which means, auctions are *anonymous*). Auctions have extensively been used in participatory sensing, and have been effective in modeling the economic interactions between the PSP and the users.

There are two types of auctions, namely *regular* and *reverse* auctions. Figure 6.a and 6.b depict the functioning scheme of a regular and reverse auction. In regular auctions, a *seller* wants to sell an item, while a group of people, called *buyers* (or *bidders*), compete to buy such item by offering a *bid*. The seller then selects the highest bidder and assigns the item to him/her. Conversely, in reverse auctions a single buyer offers a reward to execute a *task* (which might be a sensing task in case of participatory sensing). Multiple sellers are then able to offer bids to execute the task. As the auction progresses, the price *decreases* as sellers compete to offer lower bids than their competitors while still meeting all of the specifications of the task. The task is then assigned to the lowest bidder(s).

Auction mechanisms are often modeled mathematically. More specifically, an auction-based mechanism computes (i) an output $o = f(b_1, b_2, \ldots, b_n)$, representing the outcome of the auction (who is the winner) and (ii) a payment vector $p_i = (p_1, p_2, \ldots p_n)$, where $p = g(b_1, b_2, \ldots, b_n)$ is the money given to the $i$ buyer (eventually, zero to every buyer except the winner).

Based on the number of objects auctioned on the market, auctions can be categorized into *single-object* and *multi-object* auctions. Among single-object auction schemes, which are commonly applied in participatory sensing systems, the most important are *first-price* and *second-price*. In first-price auctions, the auctioneer grants the item to the highest bidder and charges the highest bid. In second-price auctions, the auctioneer grants the item to the highest bidder, but charges the second highest bid. Another type of auction is *all-pay* auction, in which every bidder pays the proposed bid regardless of whether they win the prize, which is awarded to the highest bidder as in a conventional auction.

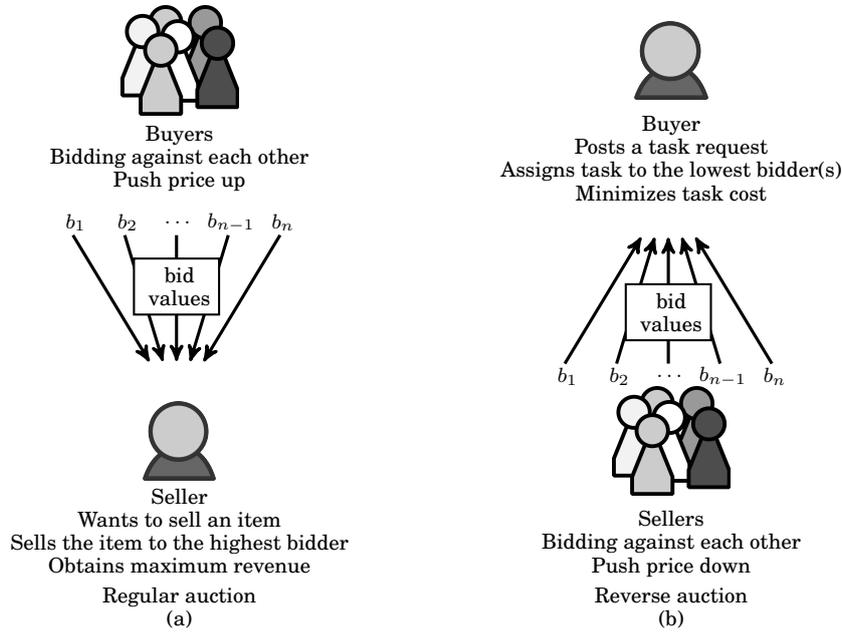

Fig. 6. Regular vs. Reverse auction.

Four characteristics should be achieved when designing an auction-based incentive mechanism:

— **Individual Rationality.** Each user (or bidder) can expect a non-negative profit.
— **Truthfulness.** Revealing true private valuation is the dominant strategy for each bidder. In other words, no bidder can improve its utility by submitting a bid different from its true valuation, no matter how others submit.
— **System Efficiency.** An auction is system-efficient if the sum of valuations of all the winning bids is maximized over the possible sets of valuations.
— **Computation Efficiency.** The outcome of the auction, which means, the set of winners among all bidders, can be computed in polynomial time.

Truthfulness of a mechanism can be guaranteed by using the following theorem.

THEOREM 1. *[Singer 2010] An auction mechanism is truthful if and only if:*

— *The selection rule is monotone: if user $i$ wins the auction by bidding $b_i$, it also wins by bidding $b'_i \leq b_i$;*
— *Each winner is paid the critical value, which means, user $i$ would not win the auction if it bids higher than this value.*

Furthermore, one of the ways to ensure truthfulness is to use the second-price *Vickrey-Clarke-Groves* auction (VCG) mechanism [Vickrey 1961]. In a forward auction, the VCG mechanism charges each individual the harm it causes to other bidders in terms of the social welfare utility, while in a reverse auction, it pays each user an amount equal to the value contributed by the user to the auction. However, it can be shown that finding the winning set of users in a VCG mechanism is NP-complete. Also, VCG auction is vulnerable to bidder collusion. In particular, if all bidders in a VCG auction reveal their valuations to each other, they can lower some or all of their valuations, while preserving who wins the auction.

### 6.2. Auction-based incentive mechanisms

Much research has been devoted to designing auction-based incentive mechanisms for participatory sensing. Table II summarizes the relevant related work on the topic in alphabetic order of the authors, with a summary of the main contribution. For each paper, the set of desirable properties achieved by the proposed auction mechanism is also reported.

Table II. Summary of auction-based incentive mechanisms. Legend: IR: Individual Rationality, T: Truthfulness, SE: System Efficiency, CE: Computation Efficiency.

| Paper | Summary of contributions | IR | T | SE | CE |
|---|---|---|---|---|---|
| [Feng et al. 2014a] | Location-aware incentive mechanism | ✓ | ✓ | ✗ | ✓ |
| [Feng et al. 2014b] | Incentive mechanisms with dynamic sensing task arrival | ✓ | ✓ | ✗ | ✓ |
| [Gao et al. 2015] | Incentive mechanisms in offline and online bidding settings | ✓ | ✓ | ✗ | ✗ |
| [Han et al. 2013] | Incentive mechanisms in offline and online bidding settings | ✓ | ✓ | ✗ | ✓ |
| [Jaimes et al. 2012] | Greedy Incentive Algorithm (GIA) based on RAPD | ✓ | ✗ | ✓ | ✓ |
| [Koutsopoulos 2013] | Optimal auction-based mechanism including QoS and QoI | ✓ | ✓ | ✗ | ✗ |
| [Lee and Hoh 2010b], [Lee and Hoh 2010a] | Reverse auction-based Dynamic Price (RAPD) mechanism | ✓ | ✗ | ✓ | ✓ |
| [Luo et al. 2014c] | All-pay auction with risk-averse users and strictly individual-rational | ✓ | ✓ | ✓ | ✓ |
| [Yang et al. 2012] | *MSensing* reverse auction-based mechanism | ✓ | ✓ | ✗ | ✓ |
| [Zhao et al. 2014], [Zhao et al. 2015] | Two online budget-feasible auction mechanisms | ✓ | ✓ | ✗ | ✓ |

The earliest work in this category is [Lee and Hoh 2010b], which was later extended in [Lee and Hoh 2010a]. This paper presents the design and evaluation of a simple incentive mechanism called Reverse Auction-based Dynamic Price (RADP). In such auction, user $u_i$ tries to sell his sensing data to a service provider by sending his claimed bid prices $b_i$. The $m$ users having lowest bid price win the auction and are subsequently rewarded with a payoff proportionate to their claimed bid price. In order to prevent users that are not rewarded from dropping out of the participatory sensing application, the authors also propose a scheme using the concept of *virtual credit* as a reward for user participation itself. In detail, a user $u_i$ who lost in the previous auction round $r-1$ receives virtual participation credit $v$ in current auction round $r$. Such credit can only be used for decreasing bid price, thus increasing winning probability of user for current auction round. Another important element of RADP is the Return on Investment (ROI) indicator $S_i$, which is used as a criterion to determine when a user is dropping out of the system. ROI index is a function of earned reward over time and the user cost of service. Users evaluate the $S_i$ value every round; if it is below a certain threshold, then they drop out of the system. In addition, RADP includes a rejoin mechanism, which allows the auctioneer to communicate the maximum winning price $\phi_k$ to the users that churned from the system.

Jaimes *et al.* pointed out in [Jaimes et al. 2012] that [Lee and Hoh 2010b] could be improved by considering the location of the users, area coverage, and budget constraints in the proposed auction-based mechanism. Therefore, they propose a novel mechanism called Greedy Incentive Algorithm (GIA), which is based on the RAPD mechanism discussed above and the Greedy Budgeted Maximum Coverage (GBMC) algorithm included in [Khuller et al. 1999]. In detail, GBMC addresses the coverage problem with budget constraints (which is shown to be NP-Hard) by greedily finding a set of users that covers the greatest possible area within a budget constraint.

The problem of guaranteeing a truthful incentive mechanism was explored for the first time in the seminal work [Yang et al. 2012]. In this paper, the authors propose a *user-centric* model for participatory sensing in which the PSP announces a set of tasks $\Gamma = \{\tau_1, \tau_2, \ldots, \tau_n\}$. Each task $\tau_j$ has a value $\nu_j > 0$ to the PSP. Each user $i$ selects a subset of tasks $\Gamma_i \subseteq \Gamma$ according to its preference. Furthermore, based on the selected task set, user $i$ also has an associated cost $c_i$, which in contrast from [Lee and Hoh 2010b], [Jaimes et al. 2012] is *private* and only known to the user. User $i$ then submits the task-bid pair $(\Gamma_i, b_i)$ to the platform, where $b_i$ is bidding price of user $i$. Upon receiving the task-bid pairs from all the users, the PSP selects a subset $\mathcal{S}$ of users as winners and determines the payment $p_i$ for each winning user $i$. The utility of user $i$ is $p_i - c_i$, while the utility of the PSP is defined as

$$u_0 = v(\mathcal{S}) - \sum_{i \in \mathcal{S}} p_i, \text{ where}$$
$$v(\mathcal{S}) = \sum_{\tau_j \in \bigcup_{i \in \mathcal{S}} \Gamma_i} \nu_j \quad (2)$$

which is, the sum of value of the tasks executed by the selected users minus the price paid to the users for execution of the sensing tasks. The proposed auction mechanism, called *MSensing*, selects winners and determines payments by following a greedy approach, therefore it is computation-efficient. This, however, makes the auction not system-efficient because the user selection and payment is not optimal. The auction is shown to be truthful by using Theorem 1, as well as individual-rational. Therefore, 4 out of 5 properties are guaranteed by *MSensing*.

In [Koutsopoulos 2013], Koutsopoulos tackles the problem of truthful incentive mechanism and proposes a mechanism based on an *optimal* reverse auction [Myerson 1981], in which the PSP needs to fulfill a given quality of service (QoS) constraint by "buying" different participation levels from different users. The novelty with respect to [Yang et al. 2012] lies in the consideration of QoS constraints, as well as the aspect of quality of information (QoI) of sensing reports in the winner determination phase. In detail, the QoI of sensing reports is modeled by an empirical *quality indicator* $q_i$ continuously updated by the PSP, which essentially measures the relevance or usefulness of information provided by user $i$ in the past. The author claims that this can be quantified by the average deviation of submitted samples from the result of the aggregation of all user samples. On the other hand, QoS is captured by a generic, positive-valued function $g(x)$ of participation level vector $x = (x_1 \cdots x_n)$, which includes as parameters the vector of qualities of submitted data by users, $q = (q_1, \cdots q_n)$. For example, in applications that consider monitoring of a quantity of interest, like air pollution or road traffic condition, $g(\cdot)$ may denote the accuracy of estimation. The target of the mechanism $M(c) = \{x(c), p(c)\}$ is then to compute a winner vector $x(c)$ and a payment level vector $p(c)$ which are function of the cost vector $c$, such that its expected expenditure for reimbursing users is minimized. A mechanism is then proposed that guarantees truthfulness and individual-rationality.

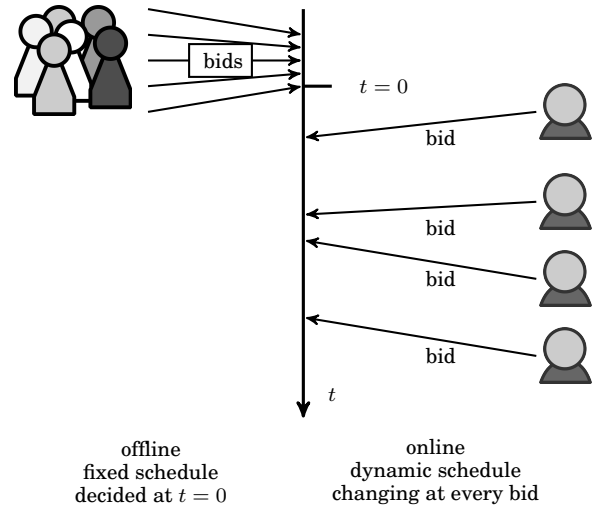

Fig. 7. Offline vs. online auction setting in [Han et al. 2013].

[Luo et al. 2014c] tackles for the first time the problem of providing a truthful mechanism which is also *system-efficient*. In particular, previous work had dealt with NP-hard scheduling problems which do not allow optimum revenue for the PSP. In this paper, the authors set the prize as a *function* of the winners' contribution in order to induce the maximum profit, which is defined as the total contribution acquired from all the users less the contribution-dependent prize. In addition, they assume users may be *risk-averse*, which means that users may be reluctant to accept a bargain with an uncertain payoff rather than another bargain with a more certain, but possibly lower, expected payoff. They then design an all-pay auction mechanism which maximizes the revenue of the PSP and is *strictly* individual-rationality for both risk-neutral and weakly risk-averse agents. That is, such agents can expect payoff strictly greater than zero and therefore are more incentivized to participate.

In [Han et al. 2013], Han *et al.* studied for the first time the problem of designing truthful incentive mechanisms for participatory sensing in an *online* bidding setting. In particular, in an offline setting, the PSP collects all the users' bids before deciding the winners and scheduling the sensing tasks. Conversely, in an online setting, the users' bids are revealed one by one, and the PSP must make an irrevocable decision on assigning sensing tasks right at the moment when the user's bid is revealed (see Figure 7). Note that this setting is also applicable to a scenario where users arrive at the sensing area at different moments in time. The authors first define the optimum task scheduling problem, which includes for the first time the possible sensing start time $s_i$ and end $e_i$ for each user. After showing the NP-hardness of the problem, they propose two polynomial-time approximation algorithms that approximate the optimum by a value $\mathcal{O}(1)$. Finally, two incentive mechanisms are proposed for the online and offline settings, which are shown to be truthful, individual-rational, and computation-efficient as they both run in $\mathcal{O}(n^2)$.

In [Zhao et al. 2014] (extended in [Zhao et al. 2015]), the problem of user incentivization in an online auction setting is further explored. In particular, the problem here is to select a subset of users before a specified deadline, so that the value of services provided by selected users is maximized under the condition that the total payment to these users does not exceed a *budget constraint*. To this end, the authors propose two online incentive mechanisms, called OMZ and OMG, which guarantee the desired properties of auctions and achieve a competitive ratio[6] of $\mathcal{O}(1)$ with respect to the offline case. In [Feng et al. 2014b], the authors also study the problem of online sensing task scheduling as in [Han et al. 2013] and [Zhao et al. 2014], by also considering *dynamic arrival* of sensing tasks. In particular, they propose two truthful auction mechanisms which take both dynamic smartphones and random arrivals of tasks into consideration. For the offline case, they design an efficient truthful mechanism with an optimal task allocation algorithm of polynomial-time computation complexity $\mathcal{O}(n+\gamma)^3$, where $n$ is the number of smartphones and $\gamma$ is the number of sensing tasks. For the online case, the authors design a near-optimal truthful mechanism with an online task allocation algorithm that achieves a constant competitive ratio of $\theta(1/2)$. It is also demonstrated that the proposed auction mechanisms achieve truthfulness, individual rationality, and computational efficiency. The same authors proposed in [Feng et al. 2014a] another incentive mechanism which takes into account the *location* of the smartphone users. Specifically, the tasks here are location-based, and users can bid only on tasks which are in the sensing coverage of the smartphone. After proving that optimally determining the winning bids with location awareness is NP-hard, the authors proposed mechanism consists of a polynomial time and near-optimal task allocation algorithm, as well as a payment scheme that guarantees truthfulness. Differently from [Feng et al. 2014b], here the sensing tasks and users' positions are assumed to be known in advance and static.

To the best of our knowledge, the most recent contribution is [Gao et al. 2015]. In this paper, the authors consider the sensor selection problem in a general time-dependent and location-aware participatory sensing system, taking the long-term user participation incentive into explicit consideration as in [Lee and Hoh 2010b]. To this end, they propose a VCG auction policy for the on-line sensor selection, which achieves a constant competitive ratio of $\mathcal{O}(1)$ with the optimal offline solution.

*6.2.1. Discussion.* Despite the merit of being the first auction-based mechanisms proposed for participatory sensing systems, [Lee and Hoh 2010b] and [Jaimes et al. 2012] do not guarantee all four properties of auctions. In particular, these approaches do not consider that cost information is private, and that users may misrepresent their real costs in order to maximize their own payoffs. As a result, the schemes are not truthful. Given GIA and RAPD algorithms determine the winners of the auction in polynomial time and guarantee non-negative utility, the mechanisms are system-efficient and individual-rational. Differently from [Lee and Hoh 2010b], [Jaimes et al. 2012] does not select the users in an optimum way due to the hardness of the coverage problem, and therefore it is not system-efficient.

Although the work in [Koutsopoulos 2013] provides the only attempt to formalize information quality as part of the incentive mechanisms, the formulation of QoS and QoI does not take into account the actual complexity of determining the quality of each sensing report. Indeed, no algorithm is proposed to accomplish such a goal. The authors also fail to calculate the complexity of the optimization problem defined in the paper, nor are algorithms formalized to solve the optimization. Therefore, the system and computation efficiency of the mechanism are not guaranteed. Furthermore, the main weakness of [Yang et al. 2012], [Feng et al. 2014a] and [Gao et al. 2015] lies in the assumption of static tasks and users. While the mathematical analysis in these papers is sound, such assumption makes the approaches unrealistic for practical applications.

---

[6]A mechanism is $\mathcal{O}(g(n))$-competitive if the ratio between the online solution and the optimal solution is $\mathcal{O}(g(n))$.

Similarly to non-auction-based mechanisms, the limitations of existing work in designing auction-based mechanism is the assumption of perfect rationality of participants. Furthermore, a large-scale validation of auction-based mechanisms is still missing, which is needed to understand whether auction-based mechanisms can work in real-world implementations.

### 6.3. Preliminaries on Game Theory

Game theory is "the study of mathematical models of conflict and cooperation between intelligent rational decision-makers" [Myerson 1991]. Originally developed to model problems in the field of economics, game theory has recently been applied to a variety of problems, in most cases, to solve the resource allocation problems in competitive environments [Hassan et al. 2012], [Birje et al. 2014], including the design of incentive mechanisms for participatory sensing systems. This is because the entities involved (i.e., the participating users) are autonomous agents, making decisions only for their own interests. To this end, game theory provides sufficient theoretical tools to analyze the users' behaviors and actions.

*6.3.1. General concepts.* Game theory models scenarios where individual decision-makers have to choose specific actions that have mutual or possibly conflicting consequences. A game consists of three major components:

— *Players:* Players are the entities invoved in the decision making process, and are denoted by a finite set $\mathcal{N} = \{1, 2, ..., n\}$.
— *Strategies:* Each player can choose a strategy $s_i$ in the set $S_i$ of possible strategies. A strategy profile $s$ is a vector containing all the strategies, i.e., $s = (s_1, s_2, \ldots, s_n)$. Obviously, we have $s \in S = \otimes_{i \in \mathcal{N}} S_i$, where $\otimes$ is the Cartesian product.
— *Utility or Payoff:* The utility $u_i : S \to \mathbb{R}$ of player $i$ is a measurement function on the possible outcome determined by the strategies of all players, where $\mathbb{R}$ is the set of real numbers.

Players are usually assumed to be rational and selfish, which means each player is only interested in maximizing its own utility without respecting others' and the system's performance [Tadelis 2013]. Let $s_{-i}$ denote any strategy profile excluding $s_i$, and $s = (s_i, s_{-i})$ for simplicity. We say that player $i$ prefers $s_i$ to $s_i'$ if $u_i(s_i, s_{-i}) > u_i(s_i', s_i)$. When other players' strategies are fixed, player $i$ can select a strategy, denoted by $b_i(s_{-i})$, which maximizes its utility function. Such a strategy is called a *best response* of player $i$. A strategy of player $i$ is *dominant* if, regardless of what other players do, the strategy earns player $i$ a larger utility than any other strategy. In order to study the interactions among players, the concept of *Nash Equilibrium* (NE) is introduced. A strategy profile constitutes an NE if none of the players can improve its utility by unilaterally deviating from its current strategy.

DEFINITION 1. *A strategy profile $s^*$ is a Nash Equibilirium (NE) if*
$$u_i(s^*) \geq u_i(s^*_{-i}, s_i') \ \forall i, \forall s_i' \in S_i$$

Games can be classified into two categories, *strategic* or *extensive* form game. The strategic form game is played only once. In this game, the players make their decisions simultaneously without knowing what others will do. On the contrary, the extensive form game represents the structure of interactions between players and defines possible orders of moves in different games. The *repeated* game is a class of the extensive form game, in which the strategic game is repeated every time. At the beginning of each stage, players observe the past history of strategies before making decisions. The number of stages may be finite or infinite. The utility of each player is the accumulated or discounted utility through all the stages. Therefore, players care not only the current utility but also the future utilities.

|  | Participant B | |
|--|--|--|
|  | Low Price | High Price |
| Participant A — Low Price | 2 / 3 | 5 / 6 |
| Participant A — High Price | 3 / 4 | 2 / 8 |

Fig. 8. An example of Stackelberg equilibrium.

*6.3.2. Stackelberg games.* A Stackelberg game is an extensive form game used to model the competition between one player, called the *leader*, and a set of players, called the *followers*. In this game, the leader takes action first and then the followers take actions. The leader knows *ex ante* that the followers observe its action and take actions accordingly. The NE in the Stackelberg game is called *Stackelberg Equilibrium* (SE).

Figure 6.3.2 depicts the payoff matrix of a Stackelberg game between two competing law firms as example. We assume that Participant A is the leader, and Participant B is the follower. The actions for both A and B are low price (LP) and high price (HP). If A plays strategy LP, B would play strategy HP, as it gives player B a utility of 5 (as opposed to a utility of 2 should B play strategy LP). This leads to a utility of 6 for A. If A plays strategy HP, B would play strategy LP, as it gives B a utility of 3 (as opposed to a utility of 2 should B play strategy HP). This leads to a utility of 4 for A. Hence A would play strategy LP, since doing so would result in a utility of 6 compared to 4 by playing strategy HP. As explained before, B would play HP if A plays LP. Therefore the Stackelberg Equilibrium of this game is (LP, HP).

The *existence* of an SE is important, since an SE strategy profile is *stable* (no player has an incentive to make a unilateral change) whereas a non-NE strategy profile is unstable. Also, the *uniqueness* of an SE allows the leader to predict the behaviors of the followers and thus, enables the leader to select the optimal strategy for itself.

### 6.4. Non-auction-based incentive mechanisms

Very few incentive mechanisms involving non-auction-based game theory have been proposed. In [Hoh et al. 2012], the authors considered the problem of trustworthy crowdsourced parking, and presented an incentive platform named *TruCentive*. The target of such platform is twofold: (i) provide incentives to stimulate contribution of high-quality parking information, and (ii) prevent malicious participants from spamming the parking service with high volume of useless data. The first aim is achieved by a simple yet effective protocol in which contributors and consumers "trade" parking information. In details,

— A contributor submits a message to *TruCentive*, receiving in turn a reward of $D$ units;
— A consumer "buys" a message for available parking spot information, depositing $N$ units;

|  | *Behavior* | |
| --- | --- | --- |
| *Parking* | Honest | Dishonest |
| Successful | $D + pX$ | $R$ |
| Unsuccessful | $R$ | $D$ |

Fig. 9. Payoff matrix for consumers, Parking vs. Behavior.

— The consumer drives to the parking spot and sends back an active confirmation of whether the parking spot information led to a successful parking in a reserved space or not. If the confirmation is successful, $R$ credits are refunded to the consumer. If successful, $X$ bonus points are credited to the contributor.

The second target is achieved by introducing a game-theoretical-formulated incentive protocol. The key idea behind the approach is that consumers who successfully parks at a traded spot can re-sell that spot through *TruCentive* if they tell the truth. In particular, if consumers lie about the success status of last parking, they cannot re-sell this parking spot to the next consumer. Therefore, reward parameters are set to ensure that the average gain of re-selling is higher than the gain by lying and receiving a refund. Figure 9 depicts the payoff matrix for consumers as a function of the behavior, honest or dishonest, and last parking status, successful or unsuccessful. In order to have (Successful, Honest) and (Unsuccessful, Honest) best strategies, $D + pX \geq R \geq D$ must hold, which means, $X \geq {1}/{p}(R - D)$. This provides a lower bound for the bonus $X$. Regarding the upper bound, the authors argue that the payment strategy must be profitable for the administration of the system, therefore, the choice of $X$ must take into account total revenue and costs of providing the service.

Duan *et al.* introduced [Duan et al. 2012] and extended[Duan et al. 2014] and analysis of different incentive mechanisms to motivate the collaboration of smartphone users on data acquisition and distributed computing applications. To collect massive amounts of sensitive data from users, they propose a game-theoretical reward mechanism based on a Stackelberg game, where the PSP can earn a fixed positive revenue only if enough smartphone users agree to participate. The authors argue that this model is sound, as an incentive mechanism is effective only when it is possible to build a large enough database to support the sensing application (e.g. live traffic maps).

Specifically, the game model proposed is a two-stage Stackelberg game under incomplete information. In the first stage, the master announces the total reward $R$ to be allocated among the required minimum number $n_0$ of participating users. In order for the master to receive revenue $V$, the number of participating users $n$ must be greater or equal to $n_0$. In the second stage, each user decides whether to join the collaboration or not by predicting other users' costs and decisions, along with expected reward. Interestingly enough, the cost $C_i$ for user $u_i$ is modeled as a function of both privacy loss and energy consumption. In the analysis, it turns out that it is better to reward users' participation efforts irrespective of the result of the collaboration. This encourages users to collaborate, and hence increases the chance of collaboration success. Then, the target of the authors is to to characterize the subgame perfect equilibria (SPE) that specify players' stable choices in all stages. As expected, the results indicate that if the master knows the users' collaboration costs $C_i$, then the selection is optimum as it can involve only users with the lowest costs.

The most important contribution pertaining to this class of incentive mechanisms was introduced in [Yang et al. 2012]. In this landmark paper, the authors consider two system models: the *platform-centric model* where the PSP provides reward shared by participating users, and the *user-centric model* where users have more control over the payment they will receive. For the platform-centric model, Yang *et al.* design an incentive mechanism using a Stackelberg game, where the PSP is the leader while the users are the followers.

It is worthwhile describing this method in detail due to its remarkable theoretical properties. In such model, there is only one sensing task, and the PSP announces a total reward $R > 0$ to incentivize users' participation to it. The sensing plan of user $u_i$ is denoted by $t_i$, and represents the time units $u_i$ is willing to provide to accomplish the sensing service, whereas $c_i$ includes the sensing cost for user $u_i$. By defining as $\mathcal{U}$ the set of all users, the utility $\overline{u_i}$ of the user $u_i$ is defined as

$$\overline{u}_i = \frac{t_i}{\sum_{j \in \mathcal{U}} t_j} \cdot R - t_i \cdot c_i \tag{3}$$

which ultimately is reward minus cost. The utility of the PSP is defined as

$$\overline{u}_0 = \lambda \log \left(1 + \sum_{i \in \mathcal{U}} \log\left(1 + t_i\right)\right) - R \tag{4}$$

where $\lambda > 1$ is a system parameter, and the inner and outer log terms quantify, respectively, the value obtained from the sensing time of user $i$ and the overall participating users. The target of the PSP is to decide the optimal value of $R$ so as to maximize Equation (4), while each user decides its sensing time $t_i$ to maximize Equation (3) for the given value of $R$. This problem is modeled as a Stackelberg game, where the PSP is the leader and the participants are the followers. It is demonstrated that a unique Stackelberg equilibrium (SE) exists, and it also shown how to compute such SE. This is an important result since the utility of the PSP is maximized, and none of the users can improve its utility by unilaterally deviating from its current strategy - hence, the strategy is stable.

*6.4.1. Discussions.* The major limitation of the work in this category is the assumption of *perfect rationality* of participants and the absence of *malicious participants*. In other words, the calculation of the equilibria holds only whether the participants willingly choose to perform the action that is supposed to be best for them. However, do they *know* the action that is best for them? Are they willing to do it, *regardless* of other factors that may influence their choice (e.g., time available to perform sensing services)? How does the system perform when participants *willingly* become malicious, even it is not the most rational choice for them? These papers do not answer these questions, and instead rely on strong assumptions that allow the authors to derive a solid and comprehensive mathematical analysis that in reality, may no longer hold.

Regarding technical limitations, the performance of the *TruCentive* solution [Hoh et al. 2012], is dependent on a good choice of the parameter $X$, which represents the bonus points that are credited for a successful and honest contribution. However, the authors did not propose any methodology to estimate or set $X$ as function of other system parameters, which limits the system's applicability in real-world applications. The two-stage Stackelburg game approach introduced in [Duan et al. 2012] assumes that the optimum strategy is found by knowing the collaboration costs of all users, which is unrealistic in practical scenarios as users may (perhaps maliciously) provide incorrect information about their costs. An even stronger assumption is applied in the platform-centric Stackelburg game introduced in [Yang et al. 2012], which assumes that costs and sensing times are perfectly known by the PSP.

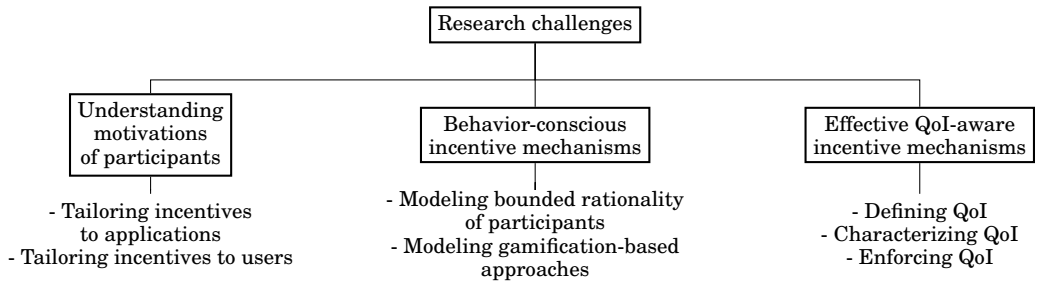

Fig. 10. Summary of proposed research challenges.

## 7. OPEN RESEARCH CHALLENGES

The last years have seen the emergence of research devoted to designing incentivization mechanisms for participatory sensing systems. While a significant number of solutions applicable to multiple contexts have been proposed, there are still some research challenges to be tackled to allow the widespread use of effective incentive mechanisms in participatory sensing. Figure 10 summarizes the proposed research agenda to improve and transforming existing incentive mechanisms.

### 7.1. What is the best way to incentivize users?

Before developing an incentive mechanism, a fundamental question must be answered: what incentive works *best* to motivate volunteers to contribute to a participatory sensing campaign?

Understanding the motivations that drive human beings to perform actions or tasks has been the object of abundant research, especially in the psychology domain [Elliot and Covington 2001], [Maehr and Meyer 1997], [Maslow 1943]. In this context, the term *motivation* has been defined as the scientific term for the representation of the reasons for our actions, our desires, and our needs [Pardee 1990]. Motivation can be divided into two different categories, known as *intrinsic* and *extrinsic* motivation.

— Intrinsic motivation can be defined as the self-desire to seek out new things and new challenges, to analyze one's capacity, to observe and to gain knowledge [Ryan and Deci 2000b]. It is driven by an interest or enjoyment in the task itself, and exists within the individual rather than relying on external pressures or a desire for reward. Intrinsic motivation is a natural motivational tendency and is a critical element in cognitive, social, and physical development. For example, students have an intrinsic motivation when they believe they have the skills to be effective in reaching their desired goals [Ryan and Deci 2000a].
— Extrinsic motivation refers to the performance of an activity in order to attain a desired outcome and it is the opposite of intrinsic motivation [Ryan and Deci 2000b]. It comes from influences outside of the individual, and is usually used to attain outcomes that a person would not get from intrinsic motivation. Common extrinsic motivations are rewards (for example money) for showing the desired behavior, and/or the threat of punishment following misbehavior. Competition is also an extrinsic motivator because it encourages the performer to win and to beat others, not simply to enjoy the intrinsic rewards of the activity [Ryan and Deci 2000a].

In motivation theory, an *incentive* is defined as something that motivates an individual to perform an action. The study of incentive structures is central to the study of all economic activities, both in terms of individual decision-making and in terms of co-operation and competition within a larger institutional structure [Armstrong and Stephens 2005]. Incentives can be classified as either *positive* or *negative*. Positive incentives seek to motivate others by promising a reward, whereas negative incentives aim to motivate others by threatening a punishment. Incentives can be to classified according to the different ways in which they motivate agents to take a particular course of action. One common taxonomy [Dalkir 2013] divides incentives into four broad classes.

— *Remunerative*. They take place when an agent can expect some form of material reward (e.g., money) in exchange for acting in a particular way.
— *Moral*. They are said to exist where a particular choice is widely regarded as "the right thing to do", or as particularly admirable, or where the failure to act in a certain way is condemned as indecent. A person acting on a moral incentive can expect a sense of self-esteem, and approval or even admiration from his community.
— *Natural*. Similarly to intrinsic motivations, natural incentive may be curiosity, mental or physical exercise, admiration, fear, anger, pain, joy, or the pursuit of truth, or the control over things in the world or people or oneself.
— *Coercive*. A person can expect a coercive incentive when the failure to act in a particular way will result in punishment being used against her by others in the community, for example, by inflicting pain in punishment, or by imprisonment, or by confiscating or destroying their possessions.

In the context of participatory sensing, it is still unknown what would be the best way to incentivize users. More specifically, although recent research has tried to understand the relation between demographics and user participation [Christin et al. 2013; Omokaro and Payton 2014], there is currently a lack of a general, empirical study on the motivations of volunteers to perform participatory sensing tasks and the effectiveness of participatory sensing incentives across different contexts. Furthermore, to date, incentives for participatory sensing have largely been applied in an *ad hoc*, "one size fits all" manner, assuming that all applications have the same requirements and people have all the same needs. Additional study is needed to develop a systematic approach to the design and selection of incentives that are *tailored* to a particular application, and *personalized* to motivate individual volunteers to perform participatory sensing tasks.

*7.1.1. Tailoring incentives to applications.* Recent work on participatory sensing has posited that the characteristics of applications, particularly with respect to the kinds of sensing tasks required, can impact the effectiveness of incentives [Rula et al. 2014]. In particular, sensing tasks may be implicit, occurring in the background with little or no user interaction (e.g., capturing location traces over the course of a predetermined time period), or may require explicit user intervention (e.g., capturing a report of the perceived conditions at a local watershed). Some sensing tasks have spatiotemporal constraints, requiring a volunteer to be at a particular location at a particular time. Sensing tasks also differ in terms of the cognitive overhead required; some sensing tasks simply require the capture of a sample from sensors on a mobile device, while others require volunteers to provide observations, interpretations, or opinions. Experimental research with large-scale deployments of participatory sensing applications can lead to a better understanding of how incentives can be tailored to the required sensing task. More broadly, the purpose of the participatory application itself may ap-peal to a potential volunteer's intrinsic motivation and impact participation. Studies in the related domains of citizen science and crowdsourcing show that intrinsic motivators often have a greater impact than remunerative rewards on increasing the quality of task performance [Nov et al. 2011; Rogstadius et al. 2011; Kaufmann et al. 2011]. In fact, offering an incentive that appeals as an extrinsic motivator may have a negative impact on an individual's intrinsic motivation, and can therefore result in reducing the overall motivation to perform a task [Gneezy and Rustichini 2000]. Understanding what motivates individuals, by using surveys/interviews, can aid in tailoring participatory sensing incentives that are effective in garnering participation.

*7.1.2. Tailoring incentives to users.* Research that explores motivation in the context of volunteerism has shown that several factors influence the motivations of an individual, including age [Carstensen et al. 1999; Wilson 2000; Okun and Schultz 2003], race [Musick et al. 2000; Morrow-Howell et al. 1990], class [Musick et al. 2000], religion [Musick et al. 2000], gender [Eisenberg 1992; Wilson and Musick 1997; Wuthnow 1995; Raddick et al. 2009] and education [Rosenthal et al. 2010]. This suggests that there is an opportunity for *personalized* incentives that appeal to the motivations of an individual.

Researchers are beginning to explore this idea within the context of participatory sensing. Work by Omokaro et al. [Omokaro and Payton 2014] takes a first step toward exploring how demographics affect the motivations of potential participatory sensing volunteers. Preliminary results from a survey of over 200 people indicate that individuals from different age groups respond similarly to incentives intended to appeal to extrinsic motivations, but displayed significant differences in their response to incentives intended to appeal to intrinsic motivations. Another study was conducted in [Christin et al. 2013], in which the authors analyze the impact of demographics, incentives, and gathering conditions on both the importance and value of privacy by means of a questionnaire-based user study with 200 anonymous participants.

While these preliminary findings show promise for this direction of research, other factors that impact an individual's motivation to contribute to participatory sensing campaigns remain to be explored. In particular, the sensing process takes time, interrupts other activities, consumes additional battery power and data traffic bandwidth. How does these issues impact on the users' willingness to participate? What is the right amount of incentive, and how much can we ask from a particular user in terms of privacy loss, time, energy and bandwidth resources? Further quantitative and qualitative study, using surveys, interviews, and participatory sensing deployments with large and diverse user populations is needed to understand the motivations of volunteers and the effectiveness of incentives to appeal to those motivations within the unique context of participatory sensing. Research efforts should also consider the lifecycle of motivation; an individual's motivation to contribute to a participatory sensing campaign may change over time, as has been found to be the case in citizen science projects [Rotman et al. 2012].

### 7.2. Towards behavior-conscious incentive mechanisms

So far, game-theoretical-based incentive mechanisms have always implicitly assumed the users are *perfectly rational*, which means, always taking the choice that maximizes their utility. However, what happens when users behave in a way that is *less than rational*? Practical experience suggests that the assumption of perfect rationality hardly holds in reality, as human beings are driven and influenced by different motivations, for example, because of emotional or personal constraints [Pita et al. 2010], or simply because remunerative incentives are not effective. Therefore, even individuals who intend to make rational choices are often bound to make *satisfying* or *necessary* (rather than maximizing or optimizing) choices in complex situations. For these reasons, the incentive mechanisms proposed present a significant gap between the theoretical modeling and the *actual* behavior of users in participatory sensing contexts. To tackle such challenge, one possible solution may be modeling interactions by using concepts from *behavioral game theory* (BGT) [Camerer 2003a]. BGT uses experimental evidence to inform mathematical models of cognitive limits, learning rules and social utility of choices [Camerer 2003b]. It has mainly three components: (i) mathematical theories of how moral obligation and vengeance affect the way people bargain and trust each other; (ii) a theory of how limits in the brain constrain the number of steps of reasoning people naturally do; and (iii) a theory of how people learn from experience to make better strategic decisions.

*7.2.1. Modeling bounded rationality of participants.* Among others, one of the powerful tools that BGT offers is the concept of *bounded rationality* [Simon 1957] [Simon 1982]. The concept stems from the fact that in general, humans make decisions under three unavoidable constraints: (i) only limited, often unreliable, information is available regarding possible alternatives and their consequences; (ii) human mind has only limited capacity to evaluate and process the information that is available; and (iii) only a limited amount of time is available to make a decision. One of the approaches [Rubinstein 1998] to model bounded-rationality behaviors is to directly and explicitly involve *procedural* aspects of decision making in the model. A procedural aspect is defined in this context as the *rationale* that leads the decision maker to a particular behavior, which might be different from the expected one (i.e., the one that maximizes user's utility). Some examples, cast in the context of participatory sensing, are:

— Users may be concerned about the *complexity* of the strategy they employ. For example, choosing the simpler strategy instead of the optimum one might be better for some users, as demonstrated in [Kahneman and Tversky 1979].
— Users might have limited ability to *understand* the game, for example, because of incomplete information at their disposal. The information may also be *asymmetrical*, which means, some users know more than others about the game being played.
— Users may be *time-constrained* and may not be able (or willing) to play the optimum strategy according to the game being played. For example, if the interaction between the user and the PSP is repeated, the PSP might play an equilibrium according to a model based on an extensive-form game. However, the equilibrium played by the users may be calculated by considering a different time-horizon, and therefore, mismatch the ones calculated by the PSP.
— In realistic contexts, participating users might not known the details of the game been played. Therefore, they need to *learn* over time the strategy that maximizes the payoff. In such context, users might have different *learning strategies*. For example, some might use a reinforcement approach [Barto 1998] (i.e., adapting the strategy as a function of current and past payoffs), while some might use anticipatory learning [Camerer et al. 2002] to reason more thoughtfully on what the other players are doing and the game being played.

Therefore, it becomes necessary to embed these and other procedural aspects while modeling incentive mechanisms. Then, we need to identify restrictions on the space of preferences that are compatible with those procedures, and then optimize based on those restrictions. One way to incorporate procedural aspects, and therefore, uncertainty in the equilibria that will be played by users, is to study the *Quantal Response Equilibria* (QRE) of the game, concept first proposed in [McKelvey and Palfrey 1995]. QRE provide a statistical framework to analyze games that incorporate realistic limitations to rational choice modeling of games. In particular, QRE allow every strategy to be played with non-zero probability, and so any strategy is possible (though not necessarily reasonable). By far, the most common specification for QRE is the *logit* equilibrium (LQRE) [McKelvey and Palfrey 1995], where player's strategies are chosen according to the probability distribution

$$\mathbb{P}\{a_i\} = \frac{exp\{\lambda \cdot \sum_{a_{-i} \in A_{-i}} \mathbb{P}\{a_{-i}\} \cdot u_i(a_i, a_{-i})\}}{\sum_{a'_i \in A_i} exp\{\lambda \cdot \sum_{a_{-i} \in A_{-i}} \mathbb{P}\{a_i\} \cdot u_i(a'_i, a_{-i})\}} \quad (5)$$

The $\lambda$ parameter is defined as the "rationality" parameter: the choice of action becomes purely random as $\lambda \to 0$, whereas the action with the higher expected payoff is chosen as $\lambda \to \infty$. Analogously to Nash equilibria, QRE can be calculated for normal-form games [McKelvey and Palfrey 1995] and extensive-form games [McKelvey and Palfrey 1998]. The value of $\lambda$ is usually chosen according to experimental data; however, such data does not exist yet to formulate QRE for games in the context of participatory sensing. Additional research is therefore needed to model the $\lambda$ parameter and therefore, obtain more realistic equilibria.

*7.2.2. Modeling gamification-based approaches.* As we have discussed in Section 5, *gamification* [Deterding et al. 2011] has been proposed as a viable and effective technique to incentivize users' participation using a natural (i.e., based on joy or entertainment) incentive type, instead of a remunerative one. Although some gamification-based incentive mechanisms have already been proposed [Kawajiri et al. 2014] [Mukherjee et al. 2014] [Ueyama et al. 2014], it is still to be demonstrated through large-scale deployments whether these mechanisms may work effectively in the context of participatory sensing. In particular, on one hand, the leaderboard-based system used by Waze [Furchgott 2010] has proven to be effective in keeping the users in the sensing loop. However, on the other hand, recent studies [Hamari et al. 2014] cast doubts on gamification, as the authors argue that gamification effects are greatly dependent on the context in which gamification is being implemented, as well as on the users using it. Therefore, additional experimental research is needed to understand and validate the conditions by which gamification is a good strategy for user incentivization in participatory sensing, and possibly, generalize the concept to a common framework that can be used by participatory sensing systems. Indeed, by far existing gamification-based have been essentially based on heuristics, and do not provide a framework to analyze and model the behavior of users in participatory sensing under gamification-based incentive. Although some modeling languages have been proposed [Herzig et al. 2013], they are general-purpose and lacking the ability to capture the complex interactions in participatory sensing. Therefore, research effort has to be put forth to identify positive and negative aspects of the approach as a function of the specific participatory sensing considered.

### 7.3. Towards effective QoI-aware incentive mechanisms

One of the greatest limitations of existing incentive mechanisms, and more generally, of participatory sensing systems, is that they are not effective in tackling the problem of verifying the actual QoI of sensing reports. In particular, the research efforts to solve such challenge can be divided in two main categories:

— *Reputation-based frameworks*. These frameworks use concepts such as information provenance (i.e., location of users) [Wang et al. 2011] [Wang et al. 2013] or consensus-based techniques [Huang et al. 2010] to determine the reliability of users and therefore filter out unreliable reports from the systems. The main issue of reputation-based systems is that they are prone to collusion-based attacks or the location-spoofing-based attacks described in Section 4.1. Therefore, such frameworks cannot be directly used in real-world participatory sensing systems.

—*QoI-aware incentive mechanisms.* Some incentive mechanisms, for example, [Koutsopoulos 2013] [Liu et al. 2011], reward the participating users based on the perceived QoI of sensing reports. Usually, the QoI of sensing reports is calculated using heuristic functions [Tham and Luo 2013], and do not measure the actual quality of the sensing report as compared to the real-world quantity being measured.

To the best of our knowledge, the only paper that addresses at once collusion- and location-based attacks is [Restuccia and Das 2014], by proposing a methodology based on mobile security agents (see Section 5) that validates and improves the QoI of sensing reports in a realistic, scalable, and secure way. However, one of the limitations of [Restuccia and Das 2014] is that the authors do not provide a detailed definition of QoI, nor do they address the problem of *how* to determine the QoI of sensing reports. More specifically, they assume that QoI can be computed by using external functions tailored to such purpose. However, this might not be the case when *multimedia* data (e.g., audio/video feed, pictures) is being collected by the participatory sensing system.

In particular, when the data is coming from physical sensors such as temperature, pressure, or light, the QoI may be defined as the accuracy of that measurement w.r.t. the real phenomenon being monitored. However, how do we define the QoI of a picture, of a sound, of a video? In such case, the specific context of the participatory sensing application may significantly influence the definition of QoI. For example, in case of noise pollution monitoring [D'Hondt et al. 2013], the maximum noise level in decibels might be sufficient to characterize QoI. On the other hand, in case of voice recognition [Lu et al. 2011] and/or applications that count the number of speakers in a room [Xu et al. 2014], precise audio measurements (in terms of portion of spectrum detected) could be needed to recognize the person and/or how many people are currently in that room. Therefore, additional research is necessary to provide different definitions of QoI that are valid for the spectrum of participatory sensing systems nowadays available.

Furthermore, although the approach based on MSAs [Restuccia and Das 2014] is very promising, it is still necessary to model, and possibly, optimize the QoI as a function of the resources (e.g., MSAs) employed. In particular, can we design a mechanism to know in advance how many resources are needed to guarantee a certain QoI of the accepted sensing reports? Can we predict the probability of accepting an unreliable sensing report, and reduce it as a function of the resources employed? Addressing these issues is extremely challenging, given the hardly predictable behavior or participants.

## 8. CONCLUSIONS

Smart devices have revolutionized our lives and the way we interact with the surrounding environment. What was technologically a dream 20 years ago has now become reality, and with more and more users embedding pervasive technologies in their daily lives, an immense number of novel applications that drastically improve people's everyday life are now possible. Among all, participatory sensing is certainly one of the most promising paradigms, as it allows to decrease dramatically infrastructure costs and obtain detailed information about the phenomenon being monitored.

In this paper, we have characterized the state-of-the-art of incentive mechanisms in participatory sensing systems, and have proposed a set of open research challenges related to such topic. First, we provided an overview of participatory sensing and identified the main components of existing incentive mechanisms. Then, we presented a taxonomy of such mechanisms, based on their purpose and their methodology, and discussed important contributions, insights, and limitations of the state of the art within each category of the taxonomy. Finally, we proposed and discussed a set of research challenges in incentivizing users in participatory sensing, which comprise guarantee-ing quality of information, modeling the behavior of users, and understanding the best reward strategy.

Future work will be dedicated to investigate the proposed research challenges. In particular, we plan to build a large-scale deployment of a participatory sensing system on our campus, and study the impact of various incentive types, both intrinsic and extrinsic, on the frequency and quality of sensing reports by participants. We also plan to design incentive mechanisms which assume bounded user rationality, and test them on practical testbeds.

**References**

Tarek Abdelzaher, Yaw Anokwa, Péter Boda, Jeff Burke, Deborah Estrin, Leonidas Guibas, Aman Kansal, Samuel Madden, and Jim Reich. 2007. Mobiscopes for Human Spaces. *IEEE Pervasive Computing* 6, 2 (April 2007), 20–29. DOI:http://dx.doi.org/10.1109/MPRV.2007.38

Andreas Albers, Ioannis Krontiris, Noboru Sonehara, and Isao Echizen. 2013. Coupons as Monetary Incentives in Participatory Sensing. In *Collaborative, Trusted and Privacy-Aware e/m-Services*, Christos Douligeris, Nineta Polemi, Athanasios Karantjias, and Winfried Lamersdorf (Eds.). IFIP Advances in Information and Communication Technology, Vol. 399. Springer Berlin Heidelberg, 226–237. DOI:http://dx.doi.org/10.1007/978-3-642-37437-1_19

Amazon. 2014. Amazon Mechanical Turk (AMT). (2014). Retrieved December 28, 2014 from https://www.mturk.com/mturk/welcome

Michael Armstrong and Tina Stephens. 2005. *A handbook of employee reward management and practice*. Kogan Page Publishers.

Kelsey D. Atherton. 2014. Israeli students spoof Waze with fake traffic jam, http://www.popsci.com/article/gadgets/israeli-students-spoof-waze-app-fake-traffic-jam. *Popular Science* (2014).

Xuan Bao and Romit Roy Choudhury. 2010. MoVi: Mobile Phone Based Video Highlights via Collaborative Sensing. In *Proceedings of the 2010 ACM International Conference on Mobile Systems, Applications, and Services (MobiSys '10)*. ACM, New York, NY, USA, 357–370. DOI:http://dx.doi.org/10.1145/1814433.1814468

Andrew G Barto. 1998. *Reinforcement learning: An introduction*. MIT press.

Mahantesh N. Birje, Sunilkumar S. Manvi, and Sajal K. Das. 2014. Reliable resources brokering scheme in wireless grids based on non-cooperative bargaining game. *Journal of Network and Computer Applications* 39, 0 (2014), 266 – 279. DOI:http://dx.doi.org/10.1016/j.jnca.2013.07.007

Jeff Burke, Deborah Estrin, Mark Hansen, Andrew Parker, Nithya Ramanathan, Sasank Reddy, and Mani B. Srivastava. 2006. Participatory sensing. In *Proceedings of the 2006 Workshop on World-Sensor-Web: Mobile Device Centric Sensor Networks and Applications (WSW '06)*. 117–134.

Colin Camerer. 2003a. *Behavioral game theory: Experiments in strategic interaction*. Princeton University Press.

Colin F. Camerer. 2003b. Behavioural studies of strategic thinking in games. *Trends in cognitive sciences* 7, 5 (2003), 225–231.

Colin F. Camerer, Teck-Hua Ho, and Juin-Kuan Chong. 2002. Sophisticated experience-weighted attraction learning and strategic teaching in repeated games. *Journal of Economic Theory* 104, 1 (2002), 137–188.

Andrew T. Campbell, Shane B. Eisenman, Nicholas D. Lane, Emiliano Miluzzo, and Ronald A. Peterson. 2006. People-centric Urban Sensing. In *Proceedings of the 2006 ACM Annual International Workshop on Wireless Internet (WICON '06)*. ACM, New York, NY, USA, Article 18. DOI:http://dx.doi.org/10.1145/1234161.1234179

Andrew T. Campbell, Shane B. Eisenman, Nicholas D. Lane, Emiliano Miluzzo, Ronald A. Peterson, Hong Lu, Xiao Zheng, Mirco Musolesi, Kristóf Fodor, and Gahng-Seop Ahn. 2008. The Rise of People-Centric Sensing. *IEEE Internet Computing* 12, 4 (July 2008), 12–21. DOI:http://dx.doi.org/10.1109/MIC.2008.90

Laura L Carstensen, Derek M Isaacowitz, and Susan T Charles. 1999. Taking time seriously: A theory of socioemotional selectivity. *American Psychologist* (1999).

Cai Chen and Yinglin Wang. 2013. SPARC: Strategy-Proof Double Auction for Mobile Participatory Sensing. In *Proceedings of the 2013 International Conference on Cloud Computing and Big Data (CLOUDCOM-ASIA '13)*. IEEE Computer Society, Washington, DC, USA, 133–140. DOI:http://dx.doi.org/10.1109/CLOUDCOM-ASIA.2013.99

Man Hon Cheung, Fen Hou, and Jianwei Huang. 2014. Participation and reporting in participatory sensing. In *Proceedings of the 2014 International Symposium on Modeling and Optimization in Mobile, Ad Hoc, and Wireless Networks (WiOpt '14)*. 357–364. DOI:http://dx.doi.org/10.1109/WIOPT.2014.6850320

Delphine Christin, Christian Buchner, and Niklas Leibecke. 2013. What's the value of your privacy? Exploring factors that influence privacy-sensitive contributions to participatory sensing applications. In *Proceedings of the 2013 IEEE Conference on Local Computer Networks Workshops (LCN Workshops)*. 918–923. DOI:http://dx.doi.org/10.1109/LCNW.2013.6758532

Delphine Christin, Andreas Reinhardt, Salil S. Kanhere, and Matthias Hollick. 2011. A survey on privacy in mobile participatory sensing applications. *Journal of Systems and Software* 84, 11 (2011), 1928 – 1946. DOI:http://dx.doi.org/10.1016/j.jss.2011.06.073 Mobile Applications: Status and Trends.

Delphine Christin, Christian RoBkopf, Matthias Hollick, Leonardo A. Martucci, and Salil S. Kanhere. 2013. IncogniSense: An anonymity-preserving reputation framework for participatory sensing applications. *Pervasive and Mobile Computing* 9, 3 (2013), 353 – 371. DOI:http://dx.doi.org/10.1016/j.pmcj.2013.01.003"

Cisco. 2014. Visual Networking Index: Global Mobile Data Traffic Forecast Update. (2014). Retrieved December 4, 2014 from http://www.cisco.com/c/en/us/solutions/service-provider/visual-networking-index-vni/index.html

Kimiz Dalkir. 2013. *Knowledge management in theory and practice*. Routledge.

Linda Deng and Landon P. Cox. 2009. LiveCompare: Grocery Bargain Hunting Through Participatory Sensing. In *Proceedings of the 2009 ACM Workshop on Mobile Computing Systems and Applications (HotMobile '09)*. ACM, New York, NY, USA, Article 4, 6 pages. DOI:http://dx.doi.org/10.1145/1514411.1514415

Sebastian Deterding, Dan Dixon, Rilla Khaled, and Lennart Nacke. 2011. From Game Design Elements to Gamefulness: Defining "Gamification". In *Proceedings of the 2011 ACM International Academic MindTrek Conference: Envisioning Future Media Environments (MindTrek '11)*. ACM, New York, NY, USA, 9–15. DOI:http://dx.doi.org/10.1145/2181037.2181040

Ellie D'Hondt, Matthias Stevens, and An Jacobs. 2013. Participatory noise mapping works! An evaluation of participatory sensing as an alternative to standard techniques for environmental monitoring. *Pervasive and Mobile Computing* 9, 5 (2013), 681 – 694. DOI:http://dx.doi.org/10.1016/j.pmcj.2012.09.002 Special issue on Pervasive Urban Applications.

Lingjie Duan, Takeshi Kubo, Koei Sugiyama, Jianwei Huang, Teruyuki Hasegawa, and Jean Walrand. 2012. Incentive mechanisms for smartphone collaboration in data acquisition and distributed computing. In *Proceedings of the 2012 IEEE International Conference on Computer Communications (INFOCOM '12)*. 1701–1709. DOI:http://dx.doi.org/10.1109/INFCOM.2012.6195541

Lingjie Duan, Takeshi Kubo, Koei Sugiyama, Jianwei Huang, Teruyuki Hasegawa, and Jean Walrand. 2014. Motivating Smartphone Collaboration in Data Acquisition and Distributed Computing. *IEEE Transactions on Mobile Computing* 13, 10 (Oct 2014), 2320–2333. DOI:http://dx.doi.org/10.1109/TMC.2014.2307327

Nancy Eisenberg. 1992. *The caring child*. Harvard University Press.

Andrew J. Elliot and Martin V. Covington. 2001. Approach and Avoidance Motivation. *Educational Psychology Review* 13, 2 (2001), 73–92. DOI:http://dx.doi.org/10.1023/A:1009009018235

Rasool Fakoor, Mayank Raj, Azade Nazi, Mario Di Francesco, and Sajal K. Das. 2012. An Integrated Cloud-based Framework for Mobile Phone Sensing. In *Proceedings of the First Edition of the MCC Workshop on Mobile Cloud Computing (MCC '12)*. ACM, New York, NY, USA, 47–52. DOI:http://dx.doi.org/10.1145/2342509.2342520

Zhenni Feng, Yanmin Zhu, Qian Zhang, L.M. Ni, and A.V. Vasilakos. 2014a. TRAC: Truthful auction for location-aware collaborative sensing in mobile crowdsourcing. In *Proceedings of the 2014 IEEE International Conference on Computer Communications (INFOCOM '14)*. 1231–1239. DOI:http://dx.doi.org/10.1109/INFOCOM.2014.6848055

Zhenni Feng, Yanmin Zhu, Qian Zhang, Hongzi Zhu, Jiadi Yu, Jian Cao, and L.M. Ni. 2014b. Towards Truthful Mechanisms for Mobile Crowdsourcing with Dynamic Smartphones. In *Proceedings of the 2014 IEEE International Conference on Distributed Computing Systems (ICDCS '14)*. 11–20. DOI:http://dx.doi.org/10.1109/ICDCS.2014.10

Roy Furchgott. 2010. New York Times. Filling in Map Gaps With Waze Games. (2010). Retrieved December 4, 2014 from http://wheels.blogs.nytimes.com/2010/05/06/filling-in-the-map-gaps-with-waze-games/

Lin Gao, Fen Hou, and Jianwei Huang. 2015. Providing Long-Term Participatory Incentive in Participatory Sensing. In *Proceedings of the 2015 IEEE International Conference on Computer Communications (INFOCOM '15)*. –.

Shravan Gaonkar, Jack Li, Romit Roy Choudhury, Landon Cox, and Al Schmidt. 2008. Micro-Blog: Sharing and Querying Content Through Mobile Phones and Social Participation. In *Proceedings of the 2008 ACM International Conference on Mobile Systems, Applications, and Services (MobiSys '08)*. ACM, New York, NY, USA, 174–186. DOI:http://dx.doi.org/10.1145/1378600.1378620

Uri Gneezy and Aldo Rustichini. 2000. Pay Enough or Don't Pay at All. *Quarterly Journal of Economics* 115, 3 (2000), 791–810.

William I. Grosky, Aman Kansal, Suman Nath, Jie Liu, and Feng Zhao. 2007. SenseWeb: An Infrastructure for Shared Sensing. *MultiMedia, IEEE* 14, 4 (Oct 2007), 8–13. DOI:http://dx.doi.org/10.1109/MMUL.2007.82

Juho Hamari, Jonna Koivisto, and Harri Sarsa. 2014. Does Gamification Work? – A Literature Review of Empirical Studies on Gamification. In *Proceedings of the 2014 Hawaii International Conference on System Sciences (HICSS '14)*. IEEE Computer Society, Washington, DC, USA, 3025–3034. DOI:http://dx.doi.org/10.1109/HICSS.2014.377

Kai Han, Chi Zhang, and Jun Luo. 2013. Truthful Scheduling Mechanisms for Powering Mobile Crowdsensing. *CoRR* abs/1308.4501 (2013). http://arxiv.org/abs/1308.4501

Jahan A. Hassan, Mahbub Hassan, Sajal K. Das, and Arthur Ramer. 2012. Managing Quality of Experience for Wireless VOIP Using Noncooperative Games. *IEEE Journal on Selected Areas in Communications* 30, 7 (August 2012), 1193–1204. DOI:http://dx.doi.org/10.1109/JSAC.2012.120805

Philippe Herzig, Kay Jugel, Christof Momm, Michael Ameling, and Alexander Schill. 2013. GaML - A Modeling Language for Gamification. In *Proceedings of the 2013 IEEE/ACM International Conference on Utility and Cloud Computing (UCC '13)*. 494–499. DOI:http://dx.doi.org/10.1109/UCC.2013.96

Baik Hoh, Tingxin Yan, D. Ganesan, K. Tracton, T. Iwuchukwu, and Juong-Sik Lee. 2012. TruCentive: A game-theoretic incentive platform for trustworthy mobile crowdsourcing parking services. In *Proceedings of the 2012 International IEEE Conference on Intelligent Transportation Systems (ITSC '12)*. 160–166. DOI:http://dx.doi.org/10.1109/ITSC.2012.6338894

Kuan Lun Huang, Salil S. Kanhere, and Wen Hu. 2010. Are You Contributing Trustworthy Data?: The Case for a Reputation System in Participatory Sensing. In *Proceedings of the 2010 ACM International Conference on Modeling, Analysis, and Simulation of Wireless and Mobile Systems (MSWIM '10)*. ACM, New York, NY, USA, 14–22. DOI:http://dx.doi.org/10.1145/1868521.1868526

Kuan Lun Huang, Salil S. Kanhere, and Wen Hu. 2014. On the need for a reputation system in mobile phone based sensing. *Ad Hoc Networks* 12, 0 (2014), 130 – 149. DOI:http://dx.doi.org/10.1016/j.adhoc.2011.12.002

Luis G. Jaimes, Idalides Vergara-Laurens, and Miguel A. Labrador. 2012. A location-based incentive mechanism for participatory sensing systems with budget constraints. In *Proceedings of the 2012 IEEE International Conference on Pervasive Computing and Communications (PerCom '12)*. 103–108. DOI:http://dx.doi.org/10.1109/PerCom.2012.6199855

Shiyu Ji and Tingting Chen. 2014. Crowdsensing incentive mechanisms for mobile systems with finite precisions. In *Proceedings of the 2014 IEEE International Conference on Communications (ICC '14)*. 2544–2549. DOI:http://dx.doi.org/10.1109/ICC.2014.6883706

Daniel Kahneman and Amos Tversky. 1979. Prospect theory: An analysis of decision under risk. *Econometrica: Journal of the Econometric Society* (1979), 263–291.

Nicholas Kaufmann, Thimo. Schulze, and Daniel Veit. 2011. More than fun and money: worker motivation in crowdsourcing–a study on Mechanical Turk. In *Proceedings of the Seventeenth Americas Conference on Information Systems, Detroit, MI*.

Ryoma Kawajiri, Masamichi Shimosaka, and Hisashi Kahima. 2014. Steered Crowdsensing: Incentive Design Towards Quality-oriented Place-centric Crowdsensing. In *Proceedings of the 2014 ACM International Joint Conference on Pervasive and Ubiquitous Computing (UbiComp '14)*. ACM, New York, NY, USA, 691–701. DOI:http://dx.doi.org/10.1145/2632048.2636064

Wazir Z. Khan, Yang Xiang, Mohammed Y. Aalsalem, and Quratulain Arshad. 2013. Mobile Phone Sensing Systems: A Survey. *IEEE Communications Surveys and Tutorials* 15, 1 (First 2013), 402–427. DOI:http://dx.doi.org/10.1109/SURV.2012.031412.00077

Samir Khuller, Anna Moss, and Joseph (Seffi) Naor. 1999. The budgeted maximum coverage problem. *Inform. Process. Lett.* 70, 1 (1999), 39 – 45. DOI:http://dx.doi.org/10.1016/S0020-0190(99)00031-9

Iordanis Koutsopoulos. 2013. Optimal incentive-driven design of participatory sensing systems. In *Proceedings of the 2013 IEEE International Conference on Computer Communications (INFOCOM '13)*. 1402–1410. DOI:http://dx.doi.org/10.1109/INFCOM.2013.6566934

Andreas Krause, Eric Horvitz, Aman Kansal, and Feng Zhao. 2008. Toward Community Sensing. In *Proceedings of the 2008 ACM/IEEE International Conference on Information Processing in Sensor Networks (IPSN '08)*. 481–492. DOI:http://dx.doi.org/10.1109/IPSN.2008.37

Vijay Krishna. 2009. *Auction theory*. Academic press.

Ioannis Krontiris, Felix C. Freiling, and Tassos Dimitriou. 2010. Location privacy in urban sensing networks: research challenges and directions [Security and Privacy in Emerging Wireless Networks]. *IEEE Wireless Communications* 17, 5 (October 2010), 30–35. DOI:http://dx.doi.org/10.1109/MWC.2010.5601955

Nicholas D. Lane, Emiliano Miluzzo, Hong Lu, Daniel Peebles, Tanzeem Choudhury, and Andrew T. Campbell. 2010. A survey of mobile phone sensing. *IEEE Communications Magazine* 48, 9 (Sept 2010), 140–150. DOI:http://dx.doi.org/10.1109/MCOM.2010.5560598

Juong-Sik Lee and Baik Hoh. 2010a. Dynamic pricing incentive for participatory sensing. *Pervasive and Mobile Computing* 6, 6 (2010), 693 – 708. DOI:http://dx.doi.org/10.1016/j.pmcj.2010.08.006 Special Issue PerCom 2010.

Juong-Sik Lee and Baik Hoh. 2010b. Sell your experiences: a market mechanism based incentive for participatory sensing. In *Proceedings of the 2010 IEEE International Conference on Pervasive Computing and Communications (PerCom '10)*. 60–68. DOI:http://dx.doi.org/10.1109/PERCOM.2010.5466993

Qinghua Li and Guohong Cao. 2013. Providing privacy-aware incentives for mobile sensing. In *Proceedings of the 2013 IEEE International Conference on Pervasive Computing and Communications (PerCom '13)*. 76–84. DOI:http://dx.doi.org/10.1109/PerCom.2013.6526717

Qinghua Li and Guohong Cao. 2014. Providing Efficient Privacy-Aware Incentives for Mobile Sensing. In *Proceedings of the 2014 IEEE International Conference on Distributed Computing Systems (ICDCS '14)*. 208–217. DOI:http://dx.doi.org/10.1109/ICDCS.2014.29

Chi H. Liu, Jun Fan, Pan Hui, Jie Wu, and Kin K. Leung. 2014. Towards QoI and Energy-Efficiency in Participatory Crowdsourcing. *IEEE Transactions on Vehicular Technology* PP, 99 (2014), 1–1. DOI:http://dx.doi.org/10.1109/TVT.2014.2367029

Chi H. Liu, Pan Hui, Joel W. Branch, Chatschik Bisdikian, and Bo Yang. 2011. Efficient network management for context-aware participatory sensing. In *Proceedings of the 2011 Annual IEEE Communications Society Conference on Sensor, Mesh and Ad Hoc Communications and Networks (SECON'11)*. 116–124. DOI:http://dx.doi.org/10.1109/SAHCN.2011.5984882

Hong Lu, A. J. Bernheim Brush, Bodhi Priyantha, Amy K. Karlson, and Jie Liu. 2011. SpeakerSense: Energy Efficient Unobtrusive Speaker Identification on Mobile Phones. In *Proceedings of the 2011 International Conference on Pervasive Computing (Pervasive '11)*. Springer-Verlag, Berlin, Heidelberg, 188–205. http://dl.acm.org/citation.cfm?id=2021975.2021992

Tie Luo, Salil S. Kanhere, Sajal K. Das, and Tan Hwee-Pink. 2014b. Optimal Prizes for All-pay Contests in Heterogeneous Crowdsensing. In *Proceedings of the 2014 IEEE International Conference on Mobile Ad-Hoc and Sensor Systems (MASS '14)*. 1–9. DOI:http://dx.doi.org/10.1109/MASS.2014.34

Tie Luo, Salil S. Kanhere, and Hwee-Pink Tan. 2014a. SEW-ing a Simple Endorsement Web to Incentivize Trustworthy Participatory Sensing. In *Proceedings of the 2014 Annual IEEE Communications Society Conference on Sensor, Mesh and Ad Hoc Communications and Networks (SECON '14)*.

Tie Luo, Hwee-Pink Tan, and Lirong Xia. 2014c. Profit-maximizing incentive for participatory sensing. In *Proceedings of the 2014 IEEE International Conference on Computer Communications (INFOCOM '14)*. 127–135. DOI:http://dx.doi.org/10.1109/INFOCOM.2014.6847932

Tie Luo and Chen-Khong Tham. 2012. Fairness and social welfare in incentivizing participatory sensing. In *Proceedings of the 2012 Annual IEEE Communications Society Conference onSensor, Mesh and Ad Hoc Communications and Networks (SECON '12)*. 425–433. DOI:http://dx.doi.org/10.1109/SECON.2012.6275807

Huadong Ma, Dong Zhao, and Peiyan Yuan. 2014. Opportunities in Mobile Crowd Sensing. *IEEE Communications Magazine* 48, 9 (Aug 2014), 29–35. DOI:http://dx.doi.org/10.1109/MCOM.2010.5560598

Martin L. Maehr and Heather A. Meyer. 1997. Understanding Motivation and Schooling: Where We've Been, Where We Are, and Where We Need to Go. *Educational Psychology Review* 9, 4 (1997), 371–409. DOI:http://dx.doi.org/10.1023/A:1024750807365

Abraham Harold Maslow. 1943. A theory of human motivation. *Psychological review* 50, 4 (1943), 370.

Richard D. McKelvey and Thomas R. Palfrey. 1995. Quantal response equilibria for normal form games. *Games and Economic Behavior* 10, 1 (1995), 6–38.

Richard D. McKelvey and Thomas R. Palfrey. 1998. Quantal response equilibria for extensive form games. *Experimental Economics* 1, 1 (1998), 9–41. DOI:http://dx.doi.org/10.1007/BF01426213

Diego Méndez, Alvaro J. Perez, Miguel A. Labrador, and Juan J. Marron. 2011. P-Sense: A participatory sensing system for air pollution monitoring and control. In *Proceedings of the 2011 IEEE International Conference on Pervasive Computing and Communications Workshops (PERCOM Workshops '11)*. 344–347. DOI:http://dx.doi.org/10.1109/PERCOMW.2011.5766902

Emiliano Miluzzo, Nicholas D. Lane, Shane B. Eisenman, and Andrew T. Campbell. 2007. CenceMe: Injecting Sensing Presence into Social Networking Applications. In *Proceedings of the 2007 European Conference on Smart Sensing and Context (EuroSSC'07)*. Springer-Verlag, Berlin, Heidelberg, 1–28. http://dl.acm.org/citation.cfm?id=1775377.1775379

Prashanth Mohan, Venakta N. Padmanabhan, and Ramachandran Ramjee. 2008. Nericell: Rich monitoring of road and traffic conditions using mobile smartphones. *Proceedings of the 2008 ACM Conference on*

*Embedded Networked Sensor Systems* (2008), 323–336. http://www.scopus.com/inward/record.url?eid=2-s2.0-84866503356&partnerID=40&md5=88b9d75452ac6f0e279bd23e039f74ef

Nancy Morrow-Howell, Leeanne Lott, and Martha Ozawa. 1990. The impact of race on volunteer helping relationships among the elderly. *Social Work* (1990).

Tridib Mukherjee, Deepthi Chander, Anirban Mondal, Koutsuv Dasgupta, Amit Kumar, and Ashwin Venkat. 2014. CityZen: A Cost-Effective City Management System with Incentive-Driven Resident Engagement. In *Proceedings of the 2014 IEEE International Conference on Mobile Data Management (MDM '14)*, Vol. 1. 289–296. DOI:http://dx.doi.org/10.1109/MDM.2014.41

Marc A Musick, John Wilson, and William B Bynum. 2000. Race and formal volunteering: The differential effects of class and religion. *Social Forces* (2000).

Roger B. Myerson. 1981. Optimal auction design. *Mathematics of operations research* 6, 1 (1981), 58–73.

Roger B. Myerson. 1991. Game theory: analysis of conflict. *Harvard University Press* (1991).

Sarfraz Nawaz and Cecilia Mascolo. 2014. Mining Users' Significant Driving Routes with Low-power Sensors. In *Proceedings of the 2014 ACM Conference on Embedded Network Sensor Systems (SenSys '14)*. ACM, New York, NY, USA, 236–250. DOI:http://dx.doi.org/10.1145/2668332.2668348

Oded Nov, Ofer Arazy, and David Anderson. 2011. Dusting for Science: Motivation and Participation of Digital Citizen Science Volunteers. In *Proceedings of the 2011 iConference (iConference)*. 68–74.

Morris A Okun and Amy Schultz. 2003. Age and motives for volunteering: testing hypotheses derived from socioemotional selectivity theory. *Psychology and aging* 18, 2 (2003), 231.

Osarieme Omokaro and Jamie Payton. 2014. *Towards a Framework to Promote User Engagement in Participatory Sensing Applications*. Technical Report. University of North Carolina at Charlotte.

Ronald L Pardee. 1990. Motivation Theories of Maslow, Herzberg, McGregor & McClelland. A Literature Review of Selected Theories Dealing with Job Satisfaction and Motivation. (1990).

James Pita, Manish Jain, Milind Tambe, Fernando Ordez, and Sarit Kraus. 2010. Robust solutions to Stackelberg games: Addressing bounded rationality and limited observations in human cognition. *Artificial Intelligence* 174, 15 (2010), 1142 – 1171. DOI:http://dx.doi.org/10.1016/j.artint.2010.07.002

M. Jordan Raddick, Georgia Bracey, Pamela L. Gay, Chris J. Lintott, Phil Murray, Kevin Schawinski, Alexander S. Szalay, and Jan Vandenberg. 2009. Galaxy Zoo: exploring the motivations of citizen science volunteers. *arXiv preprint arXiv:0909.2925* (2009).

Sasank Reddy, Deborah Estrin, Mark Hansen, and Mani Srivastava. 2010b. Examining Micropayments for Participatory Sensing Data Collections. In *Proceedings of the 2010 ACM International Conference on Ubiquitous Computing (Ubicomp '10)*. ACM, New York, NY, USA, 33–36. DOI:http://dx.doi.org/10.1145/1864349.1864355

Sasank Reddy, Deborah Estrin, and Mani Srivastava. 2010a. Recruitment Framework for Participatory Sensing Data Collections. In *Proceedings of the 2010 International Conference on Pervasive Computing (Pervasive'10)*. Springer-Verlag, Berlin, Heidelberg, 138–155. DOI:http://dx.doi.org/10.1007/978-3-642-12654-3_9

Ju Ren, Yaoxue Zhang, Kuan Zhang, and Xuemin Shen. 2014. SACRM: Social Aware Crowdsourcing with Reputation Management in Mobile Sensing. *CoRR* abs/1411.7416 (2014). http://arxiv.org/abs/1411.7416

Francesco Restuccia and Sajal K. Das. 2014. FIDES: A trust-based framework for secure user incentivization in participatory sensing. In *Proceedings of the 2014 IEEE International Symposium on A World of Wireless, Mobile and Multimedia Networks (WoWMoM '14)*. 1–10. DOI:http://dx.doi.org/10.1109/WoWMoM.2014.6918972

Jakob Rogstadius, Vassilis Kostakos, Aniket Kittur, Boris Smus, Jim Laredo, and Maja Vukovic. 2011. An Assessment of Intrinsic and Extrinsic Motivation on Task Performance in Crowdsourcing Markets. In *Proc. of AAAI Conference on Weblogs and Social Media*.

Saul Rosenthal, Candice Feiring, and Michael Lewis. 2010. Political volunteering from late adolescence to young adulthood. *Journal of Social Issues* (2010).

Dana Rotman, Jenny Preece, Jen Hammock, Kezee Procita, Derek Hansen, Cynthia Parr, Darcy Lewis, and David Jacobs. 2012. Dynamic Changes in Motivation in Collaborative Citizen-science Projects. In *Proceedings of the ACM Conference on Computer Supported Cooperative Work (CSCW)*. 217–226.

Ariel Rubinstein. 1998. Modeling Bounded Rationality. *MIT Press Books* 1 (1998).

John P. Rula, Vishnu Navda, Fabián E. Bustamante, Ranjita Bhagwan, and Saikat Guha. 2014. No "One-size Fits All": Towards a Principled Approach for Incentives in Mobile Crowdsourcing. In *Proceedings of the 15th Workshop on Mobile Computing Systems and Applications (HotMobile)*. 3:1–3:5.

Richard M. Ryan and Edward L. Deci. 2000a. Intrinsic and extrinsic motivations: Classic definitions and new directions. *Contemporary educational psychology* 25, 1 (2000), 54–67.

Richard M. Ryan and Edward L. Deci. 2000b. Self-determination theory and the facilitation of intrinsic motivation, social development, and well-being. *American psychologist* 55, 1 (2000), 68.

Herbert A Simon. 1957. Models of man; social and rational. (1957).

Herbert Alexander Simon. 1982. *Models of bounded rationality: Empirically grounded economic reason*. Vol. 3. MIT press.

Yaron Singer. 2010. Budget Feasible Mechanisms. In *Proceedings of the 2010 IEEE Annual Symposium on Foundations of Computer Science (FOCS '10)*. IEEE Computer Society, Washington, DC, USA, 765–774. DOI:http://dx.doi.org/10.1109/FOCS.2010.78

Zheng Song, Chi H. Liu, Jie Wu, Jian Ma, and Wendong Wang. 2014. QoI-Aware Multitask-Oriented Dynamic Participant Selection With Budget Constraints. *IEEE Transactions on Vehicular Technology* 63, 9 (Nov 2014), 4618–4632. DOI:http://dx.doi.org/10.1109/TVT.2014.2317701

Steven Tadelis. 2013. *Game Theory: An Introduction*. Princeton University Press.

Chen-Khong Tham and Tie Luo. 2013. Quality of Contributed Service and Market Equilibrium for Participatory Sensing. In *Proceedings of the 2013 IEEE International Conference on Distributed Computing in Sensor Systems (DCOSS '13)*. 133–140. DOI:http://dx.doi.org/10.1109/DCOSS.2013.54

Chen-Khong Tham and Tie Luo. 2014. Fairness and social welfare in service allocation schemes for participatory sensing. *Computer Networks* 73, 0 (2014), 58 – 71. DOI:http://dx.doi.org/10.1016/j.comnet.2014.07.013

Chen-Khong Tham and Tie Luo. 2015. Quality of Contributed Service and Market Equilibrium for Participatory Sensing. *IEEE Transactions on Mobile Computing* 14, 4 (April 2015), 829–842. DOI:http://dx.doi.org/10.1109/TMC.2014.2330302

Thermodo. 2014. Thermodo temperature sensor. (2014). Retrieved December 4, 2014 from http://www.thermodo.com

Arvind Thiagarajan, Lenin Ravindranath, Katrina LaCurts, Samuel Madden, Hari Balakrishnan, Sivan Toledo, and Jakob Eriksson. 2009. VTrack: Accurate, Energy-aware Road Traffic Delay Estimation Using Mobile Phones. In *Proceedings of the 2009 ACM Conference on Embedded Networked Sensor Systems (SenSys '09)*. ACM, New York, NY, USA, 85–98. DOI:http://dx.doi.org/10.1145/1644038.1644048

Brian Tung and Leonard Kleinrock. 1993. Distributed control methods. In *Proceedings of the 1993 International Symposium on High Performance Distributed Computing*. 206–215. DOI:http://dx.doi.org/10.1109/HPDC.1993.263840

Yoshitaka Ueyama, Morihiko Tamai, Yutaka Arakawa, and Keiichi Yasumoto. 2014. Gamification-based incentive mechanism for participatory sensing. In *Proceedings of the 2014 IEEE International Conference on Pervasive Computing and Communications Workshops (PERCOM Workshops '14)*. 98–103. DOI:http://dx.doi.org/10.1109/PerComW.2014.6815172

William Vickrey. 1961. Counterspeculation, auctions, and competitive sealed tenders. *The Journal of finance* 16, 1 (1961), 8–37.

Xinlei Wang, Wei Cheng, Prasant Mohapatra, and Tarek Abdelzaher. 2013. ARTSense: Anonymous reputation and trust in participatory sensing. In *Proceedings of the 2013 IEEE International Conference of Computer Communications (INFOCOM '13)*. 2517–2525. DOI:http://dx.doi.org/10.1109/INFCOM.2013.6567058

Xinlei Wang, Wei Cheng, Prasant Mohapatra, and Tarek Abdelzaher. 2014. Enabling Reputation and Trust in Privacy-Preserving Mobile Sensing. *IEEE Transactions on Mobile Computing* 13, 12 (Dec 2014), 2777–2790. DOI:http://dx.doi.org/10.1109/TMC.2013.150

Xinlei Wang, Kannan Govindan, and Prasant Mohapatra. 2011. Collusion-resilient quality of information evaluation based on information provenance. In *Proceedings of the 2011 Annual IEEE Communications Society Conference on Sensor, Mesh and Ad Hoc Communications and Networks (SECON '11)*. 395–403. DOI:http://dx.doi.org/10.1109/SAHCN.2011.5984923

Waze. 2014. The Waze Traffic Monitoring Application. (2014). Retrieved December 16, 2014 from http://www.waze.com

John Wilson. 2000. Volunteering. *Annual review of sociology* (2000).

John Wilson and Marc Musick. 1997. Who cares? Toward an integrated theory of volunteer work. *American Sociological Review* (1997).

Robert Wuthnow. 1995. *Learning to care: Elementary kindness in an age of indifference*. Oxford University Press, USA.

Chenren Xu, Sugang Li, Yanyong Zhang, Emiliano. Miluzzo, and Yi farn Chen. 2014. Crowdsensing the speaker count in the wild: implications and applications. *IEEE Communications Magazine* 52, 10 (October 2014), 92–99. DOI:http://dx.doi.org/10.1109/MCOM.2014.6917408

Yahoo! 2012. Waze sale signals new growth for Israeli high tech. (2012). Retrieved December 4, 2014 from http://news.yahoo.com/waze-sale-signals-growth-israeli-high-tech-174533585.html

Tingxin Yan, Baik Hoh, Deepak Ganesan, Kenneth Tracton, Toch Iwuchukwu, and Juong-sik Lee. 2011. CrowdPark: A Crowdsourcing-based Parking Reservation System for Mobile Phones. (2011).

Dejun Yang, Guoliang Xue, Xi Fang, and Jian Tang. 2012. Crowdsourcing to Smartphones: Incentive Mechanism Design for Mobile Phone Sensing. In *Proceedings of the 2012 ACM Annual International Conference on Mobile Computing and Networking (Mobicom '12)*. ACM, New York, NY, USA, 173–184. DOI:http://dx.doi.org/10.1145/2348543.2348567

Daqing Zhang, Haoyi Xiong, Leye Wang, and Guanling Chen. 2014. CrowdRecruiter: Selecting Participants for Piggyback Crowdsensing Under Probabilistic Coverage Constraint. In *Proceedings of the 2014 ACM International Joint Conference on Pervasive and Ubiquitous Computing (UbiComp '14)*. ACM, New York, NY, USA, 703–714. DOI:http://dx.doi.org/10.1145/2632048.2632059

Dong Zhao, Xiang-Yang Li, and Huadong Ma. 2014. How to crowdsource tasks truthfully without sacrificing utility: Online incentive mechanisms with budget constraint. In *Proceedings of the 2014 IEEE International Conference on Computer Communications (INFOCOM '14)*. 1213–1221. DOI:http://dx.doi.org/10.1109/INFOCOM.2014.6848053

Dong Zhao, Xiang-Yang. Li, and Huadong Ma. 2015. Budget-Feasible Online Incentive Mechanisms for Crowdsourcing Tasks Truthfully. *IEEE/ACM Transactions on Networking* PP, 99 (2015), 1–1. DOI:http://dx.doi.org/10.1109/TNET.2014.2379281

Pengfei Zhou, Yuanqing Zheng, and Mo Li. 2014. How Long to Wait? Predicting Bus Arrival Time With Mobile Phone Based Participatory Sensing. *IEEE Transactions on Mobile Computing* 13, 6 (June 2014), 1228–1241. DOI:http://dx.doi.org/10.1109/TMC.2013.136